\documentclass[journal, one column]{IEEEtran}
\usepackage{amsmath, amsthm, amsfonts}
\usepackage[english]{babel}
\usepackage[dvips]{graphicx}
\usepackage{graphicx}
\usepackage{tabularx}
\usepackage{enumerate}
\usepackage{mathtools}
\usepackage[ruled]{algorithm2e}
\usepackage{algorithmic}
\usepackage{tikz}
\usepackage{caption}
\usepackage{subcaption}
\newcommand{\pgftextcircled}[1]{
    \setbox0=\hbox{#1}%
    \dimen0\wd0%
    \divide\dimen0 by 2%
    \begin{tikzpicture}[baseline=(a.base)]%
        \useasboundingbox (-\the\dimen0,0pt) rectangle (\the\dimen0,1pt);
        \node[circle,draw,outer sep=0pt,inner sep=0.1ex] (a) {#1};
    \end{tikzpicture}
}

\usepackage{setspace}
\begin{document}
\title{ Distributed Multi-task APA over Adaptive Networks Based on Partial Diffusion}


\author{{Vinay Chakravarthi Gogineni$^1$, Mrityunjoy Chakraborty$^2$}\\
Department of Electronics and Electrical Communication Engineering\\
Indian Institute of Technology, Kharagpur, INDIA\\
Phone: $+91-3222-283512 \hspace{2em}$ Fax: $+91-3222-255303$\\
E.Mail : $^1\;$vinaychakravarthi@ece.iitkgp.ernet.in, $^2\;$mrityun@ece.iitkgp.ernet.in}

\maketitle
\thispagestyle{empty}

\begin{abstract}
Distributed multi-task adaptive strategies are useful to estimate multiple parameter vectors simultaneously in a collaborative manner. The existed distributed multi-task strategies use diffusion mode of cooperation in which during adaptation step each node gets the cooperation from it neighborhood nodes but not in the same cluster and during combining step each node combines the intermediate estimates of it neighboring nodes that belong to the same cluster. For this the nodes need to transmit the intermediate estimates to its neighborhood. In this paper we propose an extension to the multi-task diffusion affine projection algorithm by allowing partial sharing of the entries of the intermediate estimates among the neighbors. The proposed algorithm is termed as multi-task Partial diffusion Affine projection Algorithm (multi-task Pd-APA) which can provide the trade-off between the communication performance and the estimation performance. The performance analysis of the proposed multi-task partial diffusion APA algorithm is studied in mean and mean square sense. Simulations were conducted to verify the analytical results.
\end{abstract}
\section{Introduction}
A group of spatially-dispersed and interconnected nodes that are capable of data-processing and learning typically constitute an adaptive network. In such networks nodes the interconnected nodes continuously learn and adapt, as well as perform the assigned tasks such as parameter estimation from observations collected by the dispersed agents. Consider a connected network consisting of $N$ nodes observing temporal data arising from different spatial sources with possibly different statistical profiles. The objective is to enable the nodes to estimate a parameter vector of interest, $w_{opt}$ from the observed data. Distributed implementations offer an attractive alternative to centralized solutions with advantages related to scalability, robustness, and decentralization. Recent results in the field can be found in $\cite{1}$-$\cite{3}$. In the literature, consensus $\cite{4}$-$\cite{6}$, the incremental $\cite{7}$-$\cite{9}$, and diffusion $\cite{10}$-$\cite{13}$ strategies are most popular propositions. Incremental techniques operate on a cyclic path that runs across all nodes, which makes them sensitive to link failures and problematic for adaptive implementations. On the other hand, diffusion strategies are particularly attractive due to their enhanced adaptation performance and wider stability ranges than consensus-based implementations $\cite{14}$.
\par
The most existing literature on distributed algorithms shows that most works focus primarily on the case where the nodes estimate a single optimum parameter vector collaboratively. We shall refer to problems of this type as \emph{single-task} problems. However, many problems of interest happen to be multi-task oriented i.e., consider the general situation where there are connected clusters of nodes, and each cluster has a parameter vector to estimate. The estimation still needs to be performed cooperatively across the network because the data across the clusters may be correlated and, therefore, cooperation across clusters can be beneficial. This concept is relevant to the context of distributed estimation and adaptation over networks. Initial investigations along these lines for the traditional diffusion strategy appear in $\cite{15}$-$\cite{19}$.
\par
The most expensive part of realizing a cooperative task over a wireless ad hoc network is usually the data communications
through radio links. However, in wireless ad hoc networks, nodes often posses limited resources in terms of computational capability and electrical power.   Therefore, it is important in practice to reduce the amount of internode communications, while maintaining the benefits of cooperation. Similar to single task diffusion strategies the existing multi-task diffusion strategies relay on the cooperation among the neighboring nodes ( i.e., exchange the intermediate estimates with their neighboring nodes). Some attempts were done in single-task case to achieve this goal by partial updating $\cite{20}$-$\cite{23}$.
\par
In this paper, we propose a multi-task diffusion-based APA algorithm for distributed estimation over adaptive networks where each
node shares a part of its intermediate estimate vector with its neighbors at each iteration. The proposed algorithm, called multi-task partial-
diffusion APA (multi-task PDAPA), can reduce the internode communications relative to the multi-task full-diffusion APA algorithm, where the
entire intermediate estimate vectors are constantly transmitted, with limited sacrifice of performance. We analyze the performance
of the multi-task PDAPA algorithm in the mean and mean-square sense. In $\cite{22}$ and $\cite{23}$, LMS and RLS (recursive least squares) based algorithms for adaptive distributed estimation using partial diffusion were proposed. However, the analysis presented here is essentially different from the above techniques due to the fundamental dissimilarities of single task and multi-task diffusion adaptive strategies.
\section{Multi-task learning and Problem Formulation}
Consider a network with $N$ nodes deployed over a certain geographical area. At every time instant $n$, every node $k$ has access to time realizations $\{ d_{k}(n), \textbf{u}_{k}(n)\}$  with $d_{k}(n)$ denoting a scalar zero mean reference signal and $\textbf{u}_{k}(n)$ is an  $L \times 1$ regression vector, $\textbf{u}_{k}(n)=[u_{k}(n), u_{k}(n-1), . . . , u_{k}(n-L+1)]^{T} $ with covariance matrix $R_{u, k}=E[\textbf{u}_{k}(n)\textbf{u}^{T}_{k}(n)]$. The data at node $k$ is assumed to be related via the linear measurement model:
\begin{equation}\label{eq2.1}
\begin{split}
d_{k}(n)=\textbf{u}^{T}_{k}(n) \hspace{0.3em} \textbf{w}^{\star}_{k} + \epsilon_{k}(n)
\end{split}
\end{equation}
where $\textbf{w}^{\star}_{k}$ is an unknown optimal parameter vector to be estimated at node $k$ and $\epsilon_{k}(n)$ is an observation noise with variance $\xi^{0}$ which is assumed to be zero mean white noise and also independent of $u_{k}(n)$ for all $k$. The Nodes are grouped into Q clusters, and each cluster estimates its own optimal parameter vector. The optimum parameter vectors to be estimated are same within the cluster, but having some similarity between neighboring clusters optimal vectors, i.e.,
    \begin{equation}\label{eq2.4}
\begin{split}
\textbf{w}^{\star}_{k} = \textbf{w}^{\star}_{\mathcal{C}_{q}}, \hspace{3em} \text{whenever} \hspace{1em} k \in \mathcal{C}_{q}\hspace{2.8em}\\
\textbf{w}^{\star}_{\mathcal{C}_{p}} \sim \textbf{w}^{\star}_{\mathcal{C}_{q}}, \hspace{3em} \text{if} \hspace{1em} \mathcal{C}_{p}, \mathcal{C}_{q} \hspace{0.5em} \text{are connected}
\end{split}
\end{equation}
where $p$ and $q$ denote two cluster indexes. We say that two clusters $\mathcal{C}_{p}$ and $\mathcal{C}_{p}$ are connected if there exists at least one edge linking a node from one cluster to a node in the other cluster.
\par
In clustered multi-task networks the nodes that are grouped into cluster estimate the same coefficient vector. Thus, consider
the cluster $\mathcal{C}(k)$ to which node $k$ belongs. Under certain settings, in order to provide independence from the input data correlation statistics, we introduce normalized updates with respect to the input regressor at each node $\textbf{u}_{k}(n)$. 
\par
Following the same line of reasoning from $\cite{10}, \cite{11}$ in the single-task case, and by following same procedure mentioned in $\cite{10}$, $\cite{25}$ the following diffusion strategy of the adapt-then-combine (ATC) for clustered multi-task Normalized LMS (NLMS) is derived in distributed manner:
\begin{equation}\label{eq3.7}
\begin{split}
\begin{cases}
\boldsymbol{\psi}_{k}(n+1) = \textbf{w}_{k}(n)+ \mu \frac{\textbf{u}_{k}(n)}{\|\varepsilon+\textbf{u}_{k}(n)\|^{2}}  [\textbf{d}_{k}(n)- \textbf{u}^{T}_{k}(n)\textbf{w}_{k}(n)] + \mu_{k} \eta \sum\limits_{l \in \mathcal{N}_{k} \setminus \mathcal{C}(k)}{} \rho_{kl} ( \textbf{w}_{l}(n)-\textbf{w}_{k}(n) )  \\
\textbf{w}_{k}(n+1) = \sum\limits_{l \in \mathcal{N}_{k} \cap \mathcal{C}(k)}{} a_{lk} \boldsymbol{\psi}_{l}(n+1)   \\
\end{cases}
\end{split}
\end{equation}
By extending the above clustered multi-task diffusion strategy to data-reuse case, we can derive the following Affine projection algorithm (APA) $\cite{26}$ based clustered multi-task diffusion strategy:
\begin{equation}\label{eq3.8}
\begin{split}
\begin{cases}
\boldsymbol{\psi}_{k}(n+1) &= \textbf{w}_{k}(n)+ \mu \textbf{U}^{T}_{k}(n)\left(\varepsilon I + \textbf{U}_{k}(n)\textbf{U}^{T}_{k}(n)\right)^{-1}  [\textbf{d}_{k}(n)- \textbf{u}_{k}(n)\textbf{w}_{k}(n)] \\
& \hspace{2em}+ \mu \eta \sum\limits_{l \in \mathcal{N}_{k} \setminus \mathcal{C}(k)}{} \rho_{kl} ( \textbf{w}_{l}(n)-\textbf{w}_{k}(n) ) \\
\textbf{w}_{k}(n+1) &= \sum\limits_{l \in \mathcal{N}_{k} \cap \mathcal{C}(k)}{} a_{lk} \boldsymbol{\psi}_{l}(n+1)   \\
\end{cases}
\end{split}
\end{equation}
where $\eta$ denotes a regularization parameter with small positive value, $\varepsilon$ is employed to avoid the inversion of a rank deficient matrix $\textbf{U}_{k}(n)\textbf{U}^{T}_{k}(n)$ and the input data matrix $\textbf{U}_{k}(n)$, desired response vector $\textbf{d}_{k}(n)$ are given as follows
\begin{equation}\label{eq3.9}
\begin{split}
\textbf{U}_{k}(n)=\left[ \begin{array}{c} \textbf{u}_{k}(n) \\ \textbf{u}_{k}(n-1) \\ \vdots \\ \textbf{u}_{k}(n-P+1)  \end{array} \right] , \hspace{2em}  \textbf{d}_{k}(n)=\left[ \begin{array}{c} d_{k}(n) \\ d_{k}(n-1) \\ \vdots \\ d_{k}(n-P+1)  \end{array} \right]
\end{split}
\end{equation}
The clustered multi-task diffusion APA algorithm is given below:
\begin{algorithm}[h!]
\caption{Multi-task diffusion APA over adaptive networks}
\begin{algorithmic}
\item Start $\textbf{w}_{k}(0)=0$ for all $k$, and repeat:
\begin{equation}\label{eq3.10}
\begin{split}
\begin{cases}
\boldsymbol{\psi}_{k}(n+1)&= \textbf{w}_{k}(n) + \mu_{k} \hspace{0.2em} \textbf{U}^{T}_{k}(n) \left( \varepsilon I + \textbf{U}_{k}(n) \textbf{U}^{T}_{k}(n) \right)^{-1} [\textbf{d}_{k}(n)- \textbf{U}_{k}(n)\textbf{w}_{k}(n)] \\ &\hspace{4.6em} + \hspace{0.2em}\mu_{k} \hspace{0.2em}\eta\hspace{0.2em}\sum\limits_{l \in \mathcal{N}_{k} \setminus \mathcal{C}(k)}{} \rho_{kl} ( \textbf{w}_{l}(n)-\textbf{w}_{k}(n) )\\
\textbf{w}_{k}(n+1)&= \sum\limits_{l \in \mathcal{N}_{k} \cap \mathcal{C}(k)}{} a_{lk} \hspace{0.3em}\boldsymbol{\psi}_{l}(n+1)
\end{cases}
\end{split}
\end{equation}
\end{algorithmic}
\end{algorithm}
\subsection{ Multi-task APA with partial diffusion adaptation}
To reduce the communication load among nodes during cooperation, partial diffusion strategy $\cite{20}$ and $\cite{21}$ that aims to transmit only a subset of coefficients ($M$ in number, $M \leq L$) of intermediate estimates from each node to its neighborhood. The selection of coefficients at node $k$ and time instant $n$ can be characterized by an $L \times L$ diagonal matrix, denoted by $\textbf{S}_{k}(n)$ that has $M$ ones and $L-M$ zeros on its diagonal. The position of ones specify the selected entries. we adopted the same diffusion strategies presented in $\cite{20}$ and $\cite{21}$.
\par
Therefore the multi-task partial diffusion APA over adaptive networks is given as follows:
\begin{algorithm}[h!]
\caption{Multi-task partial diffusion APA over adaptive networks}
\begin{algorithmic}
\item Start $\textbf{w}_{k}(0)=0$ for all $k$, and repeat:
\begin{equation}\label{eq3.11}
\begin{split}
\begin{cases}
\boldsymbol{\psi}_{k}(n+1)&= \textbf{w}_{k}(n) + \mu_{k} \hspace{0.2em} \textbf{U}^{T}_{k}(n) \left( \varepsilon I + \textbf{U}_{k}(n) \textbf{U}^{T}_{k}(n) \right)^{-1} [\textbf{d}_{k}(n)- \textbf{U}_{k}(n)\textbf{w}_{k}(n)] \\ &\hspace{4.6em} + \hspace{0.2em}\mu_{k} \hspace{0.2em}\eta\hspace{0.2em}\sum\limits_{l \in \mathcal{N}_{k} \setminus \mathcal{C}(k)}{} \rho_{kl} \textbf{A}_{l}(n) \big[ \textbf{w}_{l}(n)-\textbf{w}_{k}(n) \big]\\
\textbf{w}_{k}(n+1)&= c_{kk} \hspace{0.2em} \boldsymbol{\psi}_{k}(n+1) + \sum\limits_{l \in \big(\mathcal{N}_{k} \cap \mathcal{C}(k)\big) \setminus \{k\}}{} c_{lk} \Big[ \textbf{A}_{l}(n)  \hspace{0.2em}\boldsymbol{\psi}_{l}(n+1) + \big( I_{L} - \textbf{A}_{l}(n)\big) \hspace{0.2em}\boldsymbol{\psi}_{k}(n+1)\Big]
\end{cases}
\end{split}
\end{equation}
\end{algorithmic}
\end{algorithm}
\section{Performance Analysis of  Multi-task APA with Partial Diffusion}
This section gives the performance of the proposed multi-task partial diffusion APA algorithm in mean and mean square sense.
\subsection{Network Global Model}
Before proceed to performance analysis, first, let us define the global representations as
\begin{equation}\label{eq4.1.1}
\begin{split}
\boldsymbol{\psi}(n)&=\text{col}\{\boldsymbol{\psi}_{1}(n), \boldsymbol{\psi}_{2}(n), \hdots, \boldsymbol{\psi}_{N}(n) \}, \hspace{1em} \textbf{w}(n)=\text{col}\{\textbf{w}_{1}(n), \textbf{w}_{2}(n), \hdots, \textbf{w}_{N}(n) \}\\
\textbf{U}(n)&=\text{diag}\{\textbf{U}_{1}(n), \textbf{U}_{2}(n), \hdots, \textbf{U}_{N}(n) \}, \hspace{1em} \textbf{d}(n)=\text{col}\{\textbf{d}_{1}(n), \textbf{d}_{2}(n), \hdots, \textbf{d}_{N}(n) \}\\
\end{split}
\end{equation}
where $\textbf{U}(n)$ is an $NP \times NL$ block diagonal matrix. The $NL \times NL$ diagonal matrices $\textbf{D}$ and $\boldsymbol{\eta}$ are defined by
\begin{equation}\label{eq4.1.2}
\begin{split}
\textbf{D}&=\text{blockdiag}\{\mu_{1}\textbf{I}_{L}, \mu_{2}\textbf{I}_{L}, \hdots, \mu_{N}\textbf{I}_{L} \}\\
\boldsymbol{\eta}&=\text{blockdiag}\{\eta_{1}\textbf{I}_{L}, \eta_{2}\textbf{I}_{L}, \hdots, \eta_{N}\textbf{I}_{L} \}
\end{split}
\end{equation}
to collect the local step-sizes and regularization parameters. From the linear model of the form $\eqref{eq2.1},$ the global model at network level is obtained as
\begin{equation}\label{eq4.1.3}
\begin{split}
\textbf{d}(n)= \textbf{U}(n) \textbf{w}^{\star} + \textbf{\emph{v}}(n)
\end{split}
\end{equation}
where $\textbf{w}^{\star}(n)$ and $\textbf{\emph{v}}(n)$ are global optimal weight and noise vectors given as follows
\begin{equation}\label{eq4.1.4}
\begin{split}
\textbf{w}^{\star}(n)&=\text{col}\{\textbf{w}^{\star}_{1}, \textbf{w}^{\star}_{2}, \hdots, \textbf{w}^{\star}_{N} \}\\
\textbf{\emph{v}}(n)&=\text{col}\{\textbf{\emph{v}}_{1}(n), \textbf{\emph{v}}_{2}(n), \hdots, \textbf{\emph{v}}_{N}(n) \}
\end{split}
\end{equation}
The analysis presented in $\cite{27}$ and $\cite{28}$ serves as the basis for this work. Using the above expressions, the global model of multi-task diffusion APA is therefore formulated as follows:
\begin{equation}\label{eq4.1.5}
\begin{split}
\boldsymbol{\psi}(n+1)&=  \Big[ \textbf{w}(n)+ \textbf{D} \hspace{0.2em} \textbf{U}^{T}(n) \big[\varepsilon \textbf{I} + \textbf{U}(n) \textbf{U}^{T}(n) \big]^{-1} \hspace{0.5em} [\textbf{d}(n)- \textbf{U}(n)\textbf{w}(n)] \hspace{0.2em} + \textbf{D} \hspace{0.1em} \boldsymbol{\eta} \hspace{0.1em} \boldsymbol{\mathcal{Q}}_{M}(n) \hspace{0.1em}\textbf{w}(n)\Big]\\
\textbf{w}(n+1) &= \boldsymbol{\mathcal{B}}(n) \hspace{0.2em} \boldsymbol{\psi}(n+1)
\end{split}
\end{equation}
where
\begin{equation*}\label{eq4.1.6}
\begin{split}
\boldsymbol{\mathcal{Q}}_{M}(n) &= \left[ \begin{array}{l} \textbf{Q}_{1, 1}(n) \hspace{1.5em} \cdots \hspace{1.1em}  \textbf{Q}_{1, N}(n)\\\hspace{1em} \vdots \hspace{7em} \vdots\\ \textbf{Q}_{N, 1}(n)\hspace{1em} \cdots \hspace{1em}  \textbf{Q}_{N, N}(n)\end{array}\right]\\
\textbf{Q}_{i, j}(n) &= \begin{cases} - \sum\limits_{l \in \mathcal{N}_{i}(n) \setminus \mathcal{C}(i)}^{} \rho_{i, l} \hspace{0.5em} \textbf{S}_{l}(n)    \hspace{ 1em} \text{if} \hspace{1em} i = j\\
\rho_{i, j} \hspace{0.5em} \textbf{S}_{j}(n)    \hspace{ 6.5em} \text{if} \hspace{1em} j \in \mathcal{N}_{i}(n) \setminus \mathcal{C}(i)\\
\textbf{O}_{L} \hspace{ 11.5em}  \hspace{1em} \text{otherwise}
\end{cases}\\
\boldsymbol{\mathcal{Q}}_{M}(n)&= \boldsymbol{\mathcal{P}} \odot  \boldsymbol{\mathcal{S}}(n) -  \Big( \boldsymbol{\mathcal{S}}(n) \hspace{0.5em}  \boldsymbol{\mathcal{P}}^{T} \Big) \odot \textbf{I}_{LN}\\
\boldsymbol{\mathcal{B}}(n) &= \left[ \begin{array}{l} \textbf{B}_{1, 1}(n) \hspace{1.5em} \cdots \hspace{1.1em}  \textbf{B}_{1, N}(n)\\\hspace{1em} \vdots \hspace{7em} \vdots\\ \textbf{B}_{N, 1}(n)\hspace{1em} \cdots \hspace{1em}  \textbf{B}_{N, N}(n)\end{array}\right]\\
\end{split}
\end{equation*}
\begin{align}\label{eq4.1.6}
\begin{split}
\textbf{B}_{i, j}(n) &= \begin{cases} \textbf{I}_{L} - \sum\limits_{l \in \Big( \mathcal{N}_{i}(n) \cap \mathcal{C}(i) \Big) \setminus \{i\}}^{} a_{l, i} \hspace{0.5em} \textbf{S}_{l}(n)    \hspace{ 4em} \text{if} \hspace{1em} i = j\\
a_{j, i} \hspace{0.5em} \textbf{S}_{j}(n)    \hspace{ 13em} \text{if} \hspace{1em} j\in \Big( \mathcal{N}_{i}(n) \cap \mathcal{C}(i) \Big) \setminus \{i\} \\
\textbf{O}_{L} \hspace{ 15.5em}  \hspace{1em} \text{otherwise}
\end{cases}\\
\boldsymbol{\mathcal{B}}(n)&= \boldsymbol{\mathcal{A}}^{T} \odot  \boldsymbol{\mathcal{S}}(n) +  \Big( \textbf{I}_{LN} - \boldsymbol{\mathcal{S}}(n)   \hspace{0.5em} \boldsymbol{\mathcal{A}} \Big) \odot \textbf{I}_{LN}\\
\end{split}
\end{align}
and
\begin{equation}
\begin{split}
\boldsymbol{\mathcal{P}} &= \textbf{P} \otimes \textbf{I}_{L}\\
\boldsymbol{\mathcal{A}} &= \textbf{A} \otimes \textbf{I}_{L} \\
\boldsymbol{\mathcal{S}}(n)&= \textbf{1}_{N} \otimes \Big[\textbf{S}_{1}(n), \textbf{S}_{2}(n), \cdots, \textbf{S}_{N}(n) \Big]
\end{split}
\end{equation}
with $\textbf{1}_{N}$ denoting a $N \times 1$ vector and $\textbf{O}_{L}$ denoting the $L \times L$ zero matrix. Now the objective is to study the performance behavior of the multi-task partial diffusion APA governed by the form $\eqref{eq4.1.5}$.
\subsection{Mean Error Behavior Analysis}
The global error vector $\textbf{e}(n)$ is related to the local error vectors $\textbf{e}_{k}(n)$ as
\begin{equation}\label{eq4.2.1}
\begin{split}
\textbf{e}(n)&=\text{col}\{\textbf{e}_{1}(n), \textbf{e}_{2}(n), \hdots, \textbf{e}_{N}(n) \}
\end{split}
\end{equation}
By denoting $\widetilde{\textbf{w}}(n)= \textbf{w}^{\star}-\textbf{w}(n)$
the global weight error vector can be rewritten as
\begin{equation}\label{eqeq4.2.2}
\begin{split}
\textbf{e}(n)=[\textbf{d}(n)- \textbf{U}(n)\textbf{w}(n)] \hspace{0.2em} \Big] = \textbf{U}(n) \widetilde{\textbf{w}}(n) + \textbf{v}(n)= \textbf{e}_{a}(n) + \textbf{\emph{v}}(n)
\end{split}
\end{equation}
where
\begin{equation}\label{eq4.2.3}
\begin{split}
\textbf{e}_{a}(n)= \textbf{U}(n) \widetilde{\textbf{w}}(n)
\end{split}
\end{equation}
Using these results the recursive update equation of global weight error vector can be written as
\begin{equation}\label{eq4.2.4}
\begin{split}
\widetilde{\textbf{w}}(n+1)&= \boldsymbol{\mathcal{B}}(n)\left[\begin{array}{l} \widetilde{\textbf{w}}(n)- \textbf{D} \hspace{0.1em} \textbf{U}^{T}(n) \big[\varepsilon \textbf{I} + \textbf{U}(n) \textbf{U}^{T}(n) \big]^{-1} \hspace{0.1em} \textbf{U}(n)\widetilde{\textbf{w}}(n) \hspace{0.1em} - \textbf{D} \hspace{0.1em} \textbf{U}^{T}(n) \big[\varepsilon \textbf{I} + \textbf{U}(n) \textbf{U}^{T}(n) \big]^{-1} \hspace{0.1em} \textbf{v}(n) \\
\hspace{2em} + \hspace{0.1em} \textbf{D} \hspace{0.1em}\boldsymbol{\eta} \hspace{0.2em} \boldsymbol{\mathcal{Q}}_{M}(n) \widetilde{\textbf{w}}(n) - \hspace{0.1em} \textbf{D}(n) \hspace{0.1em}\boldsymbol{\eta} \hspace{0.2em} \boldsymbol{\mathcal{Q}}_{M}(n) \textbf{w}^{\star}\end{array}\right]\\
&= \boldsymbol{\mathcal{B}}(n) \Big[ \textbf{I}_{LN} - \textbf{D} \hspace{0.2em} \textbf{U}^{T}(n) \big[\varepsilon \textbf{I} + \textbf{U}(n) \textbf{U}^{T}(n) \big]^{-1} \hspace{0.5em} \textbf{U}(n) \hspace{0.2em} + \hspace{0.2em} \textbf{D} \hspace{0.2em}\boldsymbol{\eta} \hspace{0.2em} \boldsymbol{\mathcal{Q}}_{M}(n) \Big]\widetilde{\textbf{w}}(n) \\ &\hspace{2em} - \boldsymbol{\mathcal{B}}(n) \hspace{0.1em} \textbf{D} \hspace{0.2em} \textbf{U}^{T}(n) \big[\varepsilon \textbf{I} + \textbf{U}(n) \textbf{U}^{T}(n) \big]^{-1} \hspace{0.1em} \textbf{v}(n)
 - \boldsymbol{\mathcal{B}}(n) \hspace{0.2em} \textbf{D} \hspace{0.2em}\boldsymbol{\eta} \hspace{0.2em}  \boldsymbol{\mathcal{Q}}_{M}(n) \textbf{w}^{\star} \\
\end{split}
\end{equation}
Taking the expectation $E[\cdot]$ of both sides, using the statistical independence of $\boldsymbol{\mathcal{B}}(n)$ and $\boldsymbol{\mathcal{Q}}_{M}(n)$ and recalling that $\textbf{\emph{v}}_{k}(n)$ is zero-mean
i.i.d and also independent of $\textbf{U}_{k}(n)$ and thus of $\textbf{w}_{k}(n)$, the network mean error vector can be written as follows:
\begin{equation}\label{eq4.2.5}
\begin{split}
E\big[\widetilde{\textbf{w}}(n+1)\big]= \overline{\boldsymbol{\mathcal{B}}} \bigg[ \textbf{I}_{LN} - \textbf{D} \hspace{0.2em} E\Big[\textbf{U}^{T}(n) \big[\varepsilon \textbf{I} + \textbf{U}(n) \textbf{U}^{T}(n) \big]^{-1} \hspace{0.5em} \textbf{U}(n) \Big]\hspace{0.2em} + \hspace{0.2em} \textbf{D} \hspace{0.2em}\boldsymbol{\eta} \hspace{0.2em} \overline{\boldsymbol{\mathcal{Q}}}_{M} \bigg] E[\widetilde{\textbf{w}}(n)] -  \overline{\boldsymbol{\mathcal{B}}} \hspace{0.2em} \textbf{D} \hspace{0.2em}\boldsymbol{\eta} \hspace{0.2em} \overline{\boldsymbol{\mathcal{Q}}}_{M} \textbf{w}^{\star}
\end{split}
\end{equation}
where
\begin{equation*}\label{eq4.2.5}
\begin{split}
\overline{\boldsymbol{\mathcal{Q}}}_{M}&=  \boldsymbol{\mathcal{P}}\odot E[\boldsymbol{\mathcal{S}}(n)] - \Big( E[\boldsymbol{\mathcal{S}}(n)]  \boldsymbol{\mathcal{P}}^{T} \Big) \odot \textbf{I}_{LN}\\
&=p \hspace{0.2em} \bigg(\boldsymbol{\mathcal{P}} \odot (\textbf{J}_{N} \otimes \textbf{I}_{L}) - \Big( (\textbf{J}_{N} \otimes \textbf{I}_{L})  \boldsymbol{\mathcal{P}}^{T} \Big) \odot I_{LN}\bigg)\\
&= p \hspace{0.2em} \Big(\boldsymbol{\mathcal{P}}  -  I_{LN}\Big)\\
&= p \hspace{0.2em} \boldsymbol{\mathcal{Q}}\\
\overline{\boldsymbol{\mathcal{B}}}&=  \boldsymbol{\mathcal{A}}^{T} \odot E[\boldsymbol{\mathcal{S}}(n)] + \Big( \textbf{I}_{LN} - E[\boldsymbol{\mathcal{S}}(n)]  \boldsymbol{\mathcal{A}} \Big) \odot \textbf{I}_{LN}\\
&=p \hspace{0.2em} \boldsymbol{\mathcal{A}}^{T} \odot (\textbf{J}_{N} \otimes \textbf{I}_{L}) + \Big( \textbf{I}_{LN} - p (\textbf{J}_{N} \otimes \textbf{I}_{L})  \boldsymbol{\mathcal{A}}\Big) \odot I_{LN}\\
&= p \hspace{0.2em} \boldsymbol{\mathcal{A}}^{T}  + (1-p) I_{LN}
\end{split}
\end{equation*}

with $\boldsymbol{\mathcal{Q}} = \Big(\boldsymbol{\mathcal{P}}  -  I_{LN}\Big)$ is the matrix that involves in multi-task full diffusion APA and $p= \frac{M}{L}$, is the probability that a particular entry is transmitted. Therefore, for any initial condition, in order to guarantee the stability of the multi-task partial diffusion APA strategy in the mean sense if, and only if, the step size $\mu_{k}$ has to be chosen to satisfy
\begin{equation}\label{eq4.2.6}
\begin{split}
 \rho\Big(\hspace{0.1em} \overline{\boldsymbol{\mathcal{B}}} \big[\textbf{I}_{LN}- \textbf{D} \hspace{0.2em}\overline{\textbf{Z}} + \textbf{D} \hspace{0.2em}\boldsymbol{\eta} \hspace{0.2em} \overline{\boldsymbol{\mathcal{Q}}}_{M} \big] \hspace{0.1em} \Big) < 1
\end{split}
\end{equation}
where $\overline{\textbf{Z}}=  \hspace{0.1em} E\Big[\textbf{U}^{T}(n) \big[\varepsilon \textbf{I} + \textbf{U}(n) \textbf{U}^{T}(n) \big]^{-1} \hspace{0.5em} \textbf{U}(n) \Big]\hspace{0.1em}$ and $\rho(\cdot)$ denotes the spectral radius of its argument. Since any induced matrix norm is lower bounded by the spectral radius, we can write the following relation in terms of block maximum norm:
\begin{equation}\label{eq4.2.7}
\begin{split}
 \rho\Big(\hspace{0.1em} \overline{\boldsymbol{\mathcal{B}}} \big[\textbf{I}_{LN}- \textbf{D} \hspace{0.2em}\overline{\textbf{Z}} + \textbf{D} \hspace{0.2em}\boldsymbol{\eta} \hspace{0.2em} \overline{\boldsymbol{\mathcal{Q}}}_{M} \big] \hspace{0.1em} \Big)  \leq \| \overline{\boldsymbol{\mathcal{B}}} \big[\textbf{I}_{LN}- \textbf{D} \hspace{0.2em}\overline{\textbf{Z}} + \textbf{D} \hspace{0.2em}\boldsymbol{\eta} \hspace{0.2em} \overline{\boldsymbol{\mathcal{Q}}}_{M} \big] \|_{b, \infty}
\end{split}
\end{equation}
\par
Now using the norm inequalities and the fact that as shown in the above the rows of $\overline{\boldsymbol{\mathcal{B}}}$ add up to unity, we have
\begin{equation}\label{eq4.2.8}
\begin{split}
\| \overline{\boldsymbol{\mathcal{B}}} \big[\textbf{I}_{LN}- \textbf{D} \hspace{0.2em}\overline{\textbf{Z}} + \textbf{D} \hspace{0.2em}\boldsymbol{\eta} \hspace{0.2em} \overline{\boldsymbol{\mathcal{Q}}}_{M} \big] \|_{b, \infty} &\leq \| \big[\textbf{I}_{LN}- \textbf{D} \hspace{0.2em}\overline{\textbf{Z}} + \textbf{D} \hspace{0.2em}\boldsymbol{\eta} \hspace{0.2em} \overline{\boldsymbol{\mathcal{Q}}}_{M} \big] \|_{b, \infty}\\
\end{split}
\end{equation}
Let $A$ be the an $L \times L$ matrix, then from Gershgorin circle theorem, we have:
\begin{equation}
 |\lambda - a_{i, i} | \leq \sum\limits_{j \neq i} |a_{i, j}|
\end{equation}
By assuming  $\overline{\mu}_{k}=\overline{\mu}$, and using the above result, a sufficient condition for $\eqref{eq4.2.8}$ to hold is to choose  $\mu_{k}$ such that
\begin{equation}\label{eq4.2.9}
\begin{split}
0 < \mu_{k} < \frac{2}{\text{max}_{k}\{\lambda_{max}(\overline{\textbf{Z}}_{k})\}+2 \eta \hspace{0.2em} p}
\end{split}
\end{equation}
where $\overline{\textbf{Z}}_{k}=E\Big[\textbf{U}_{k}^{T}(n) \big[\varepsilon \textbf{I} + \textbf{U}_{k}(n) \textbf{U}_{k}^{T}(n) \big]^{-1} \hspace{0.5em} \textbf{U}_{k}(n) \Big]$.
In general $0 \leq p \leq 1$. Above result clearly shows that the mean stability limit of the multi-task partial diffusion APA is lower than the diffusion APA due to the presence of $\eta$ however, it is better than the multi-task full diffusion APA. It is easy to verify that when $p=0$ that means no cooperation during adaptation and combining, the conditio on $\mu_{k}$ is same as APA (recall diffusion APA bound also same). on the other hand when $p=1$ that means full cooperation i.e., all the coefficients are sharing to neighborhood in each iteration, the limit on $\mu_{k}$ is simply becomes same as multi-task diffusion APA.
\par
In steady-state i.e., as $n \rightarrow \infty $  the asymptotic mean bias is given by
\begin{equation}\label{eq4.2.10}
\begin{split}
\lim\limits_{n \to \infty}E\big[\widetilde{\textbf{w}}(n)\big]= \Bigg[ \overline{\boldsymbol{\mathcal{B}}}  \bigg[ \textbf{I}_{LN} - \textbf{D} \hspace{0.2em} \textbf{Z} \hspace{0.2em} + \hspace{0.2em} \textbf{D} \hspace{0.2em} \boldsymbol{\eta} \hspace{0.2em} \overline{\boldsymbol{\mathcal{Q}}}_{M}   \bigg] - \textbf{I}_{LN} \Bigg]^{-1}  \overline{\boldsymbol{\mathcal{B}}}  \hspace{0.3em} \textbf{D} \hspace{0.2em} \boldsymbol{\eta} \hspace{0.2em} \overline{\boldsymbol{\mathcal{Q}}}_{M} \hspace{0.2em} \textbf{w}^{\star}
\end{split}
\end{equation}
\subsection{Mean-Square Error Behavior Analysis}
The recursive update equation of weight error vector can be rewritten as
\begin{equation}\label{eq4.3.1}
\begin{split}
\widetilde{\textbf{w}}(n+1)&= \textbf{G}(n) \widetilde{\textbf{w}}(n) - \boldsymbol{\mathcal{B}}(n) \hspace{0.1em} \textbf{D} \hspace{0.1em}\textbf{U}^{T}(n) \big[\varepsilon \textbf{I} + \textbf{U}(n) \textbf{U}^{T}(n) \big]^{-1} \textbf{\emph{v}}(n) - \textbf{r}(n)
\end{split}
\end{equation}
where
\begin{equation}\label{eq4.3.2}
\begin{split}
\textbf{G}(n)&= \boldsymbol{\mathcal{B}}(n) \Big[\textbf{I}_{LN} -  \textbf{D} \hspace{0.1em} \textbf{U}^{T}(n) \big[\varepsilon \textbf{I} + \textbf{U}(n) \textbf{U}^{T}(n) \big]^{-1} \textbf{U}(n) + \textbf{D} \hspace{0.1em} \boldsymbol{\eta} \hspace{0.1em}\boldsymbol{\mathcal{Q}}_{M}(n)\Big]\\
\textbf{r}(n)&=\boldsymbol{\mathcal{B}}(n)\hspace{0.1em}\textbf{D} \hspace{0.2em} \boldsymbol{\eta} \hspace{0.1em}\boldsymbol{\mathcal{Q}}_{M}(n)\hspace{0.1em}\textbf{w}^{\star}
\end{split}
\end{equation}
Using the standard independent assumption between $\textbf{U}_{k}(n)$ and $\textbf{w}_{k}(n)$ and $E[\emph{v}(n)]=0$, the mean square of the weight error vector $\widetilde{\textbf{w}}(n+1)$, weighted by any positive semi-definite matrix $\boldsymbol{\Sigma}$ that we are free to choose, satisfies the following relation:
\begin{equation}\label{eq4.3.3}
\begin{split}
E\|\widetilde{\textbf{w}}(n+1)\|_{\boldsymbol{\Sigma}}^{2}&= E\|\widetilde{\textbf{w}}(n)\|_{E \boldsymbol{\Sigma}^{'}}^{2}
+ E\big[\textbf{\emph{v}}^{T}(n) \hspace{0.1em}\textbf{Y}^{\boldsymbol{\Sigma}}(n) \hspace{0.1em} \textbf{\emph{v}}(n) \big] - E\Big[ \widetilde{\textbf{w}}^{T}(n) \hspace{0.2em} \textbf{G}^{T}(n) \hspace{0.1em} \boldsymbol{\Sigma} \hspace{0.3em}\textbf{r}(n) \Big]- E\Big[ \textbf{r}^{T}(n) \hspace{0.3em}\boldsymbol{\Sigma}\hspace{0.1em} \textbf{G}(n)\hspace{0.1em} \widetilde{\textbf{w}}(n) \Big] + E\|\textbf{r}(n) \|_{\boldsymbol{\Sigma}}^{2}
\end{split}
\end{equation}
where
\begin{equation}\label{eq4.3.4}
\begin{split}
E\boldsymbol{\Sigma}^{'}&= E \Big[\textbf{G}^{T}(n) \hspace{0.1em} \boldsymbol{\Sigma} \hspace{0.1em} \textbf{G}(n) \Big]\\
&=\hspace{0.5em}E \big[\boldsymbol{\mathcal{B}}^{T}(n)\hspace{0.1em} \boldsymbol{\Sigma} \hspace{0.1em} \boldsymbol{\mathcal{B}}(n) \big]
- E \big[\boldsymbol{\mathcal{B}}^{T}(n)\hspace{0.1em} \boldsymbol{\Sigma} \hspace{0.1em} \boldsymbol{\mathcal{B}}(n) \big]\hspace{0.1em} \textbf{D}  \hspace{0.1em}\overline{\textbf{Z}}
+ E \big[\boldsymbol{\mathcal{B}}^{T}(n)\hspace{0.1em} \boldsymbol{\Sigma} \hspace{0.1em} \boldsymbol{\mathcal{B}}(n) \big] \hspace{0.1em} \textbf{D} \hspace{0.1em} \boldsymbol{\eta} \overline{\boldsymbol{\mathcal{Q}}}_{M}
- \overline{\textbf{Z}}^{T}  \textbf{D}  E \big[\boldsymbol{\mathcal{B}}^{T}(n)\hspace{0.1em} \boldsymbol{\Sigma} \hspace{0.1em} \boldsymbol{\mathcal{B}}(n) \big] \\
&\hspace{1em} + E\Big[ \textbf{Z}^{T}(n) \hspace{0.2em} \textbf{D} \hspace{0.2em} \boldsymbol{\mathcal{B}}^{T}(n) \hspace{0.1em} \boldsymbol{\Sigma} \hspace{0.1em} \boldsymbol{\mathcal{B}}(n) \hspace{0.2em} \textbf{D} \hspace{0.2em} \textbf{Z}(n)\Big]
- E\Big[ \textbf{Z}^{T}(n) \hspace{0.2em} \textbf{D} \hspace{0.2em} \boldsymbol{\mathcal{B}}^{T}(n) \hspace{0.1em} \boldsymbol{\Sigma} \hspace{0.1em} \boldsymbol{\mathcal{B}}(n) \hspace{0.2em} \textbf{D} \boldsymbol{\eta}\hspace{0.2em} \boldsymbol{\mathcal{Q}}_{M}(n)\Big] \\
&\hspace{1em}+ E\Big[\boldsymbol{\mathcal{Q}}_{M}^{T}(n) \hspace{0.2em} \boldsymbol{\eta} \hspace{0.2em} \textbf{D} \boldsymbol{\mathcal{B}}^{T}(n)\hspace{0.1em} \boldsymbol{\Sigma} \hspace{0.1em} \boldsymbol{\mathcal{B}}(n) \Big]
- E\Big[\boldsymbol{\mathcal{Q}}_{M}^{T}(n) \hspace{0.2em} \boldsymbol{\eta} \hspace{0.2em} \textbf{D} \boldsymbol{\mathcal{B}}^{T}(n)\hspace{0.1em} \boldsymbol{\Sigma} \hspace{0.1em} \boldsymbol{\mathcal{B}}(n) \textbf{D} \textbf{Z}(n)\Big] \\
& \hspace{1em}+ E\Big[\boldsymbol{\mathcal{Q}}_{M}^{T}(n) \hspace{0.2em} \boldsymbol{\eta} \hspace{0.2em} \textbf{D} \boldsymbol{\mathcal{B}}^{T}(n)\hspace{0.1em} \boldsymbol{\Sigma} \hspace{0.1em} \boldsymbol{\mathcal{B}}(n) \textbf{D}  \boldsymbol{\eta}  \hspace{0.2em} \boldsymbol{\mathcal{Q}}_{M}(n)\Big]
\end{split}
\end{equation}
and
\begin{equation}\label{eq4.3.5}
\begin{split}
\textbf{Y}^{\boldsymbol{\Sigma}}(n)&=\big[\varepsilon \textbf{I} + \textbf{U}(n) \textbf{U}^{T}(n) \big]^{-1}\textbf{U}(n) \textbf{D} \hspace{0.1em} \boldsymbol{\mathcal{B}}^{T}(n) \hspace{0.1em} \boldsymbol{\Sigma} \boldsymbol{\mathcal{B}}(n) \hspace{0.1em}\textbf{D} \hspace{0.1em}    \textbf{U}^{T}(n) \big[\varepsilon \textbf{I} + \textbf{U}(n) \textbf{U}^{T}(n) \big]^{-1}\\
\end{split}
\end{equation}
To extract the matrix $\boldsymbol{\Sigma}$ from the expectation terms, a weighted variance relation is introduced by using $L^{2}N^{2} \times 1$ column vectors:
\begin{equation}\label{eq4.3.7}
\begin{split}
\boldsymbol{\sigma}= \text{bvec}\{\boldsymbol{\Sigma}\} \hspace{2em} \text{and} \hspace{2em} \boldsymbol{\sigma^{'}}= \text{bvec}\{E\boldsymbol{\Sigma}^{'}\}
\end{split}
\end{equation}
where $\text{bvec}\{\cdot\}$ denotes the block vector operator. In addition, $\text{bvec}\{\cdot\}$ is also used to recover the original matrix $\Sigma$ from $\boldsymbol{\sigma}$. One property of the $\text{bvec}\{\otimes_{b}\}$ operator when working with
the block Kronecker product $\cite{29}$ is used in this work, namely,
\begin{equation}\label{eq4.3.8}
\begin{split}
\text{bvec}\{\textbf{Q} \boldsymbol{\Sigma} \textbf{P}^{T}\}= (\textbf{P} \otimes_{b} \textbf{Q}) \boldsymbol{\sigma}
\end{split}
\end{equation}
where $\textbf{P} \otimes_{b} \textbf{Q}$ denotes the block Kronecker product $\cite{29}$, $\cite{30}$ of two block matrices.
\par
Using $\eqref{eq4.3.8}$ to $\eqref{eq4.3.4}$ after block vectorization, the following terms on the
right side of $\eqref{eq4.3.4}$ are given by
\begin{equation}\label{eq4.3.9}
\begin{split}
\text{bvec}\Big\{ E \big[\boldsymbol{\mathcal{B}}^{T}(n)\hspace{0.1em} \boldsymbol{\Sigma} \hspace{0.1em} \boldsymbol{\mathcal{B}}(n) \big] \Big\}&= E\Big( \boldsymbol{\mathcal{B}}^{T}(n) \otimes_{b} \boldsymbol{\mathcal{B}}^{T}(n) \Big) \hspace{0.2em} \boldsymbol{\sigma}
\end{split}
\end{equation}
\begin{equation}\label{eq4.3.10}
\begin{split}
\text{bvec}\Big\{E \big[\boldsymbol{\mathcal{B}}^{T}(n)\hspace{0.1em} \boldsymbol{\Sigma} \hspace{0.1em} \boldsymbol{\mathcal{B}}(n) \big]\hspace{0.1em} \textbf{D}  \hspace{0.1em}\overline{\textbf{Z}} \Big\}&=  \big( \overline{\textbf{Z}} \otimes_{b} \textbf{I}_{LN} \big)  \big(  \textbf{D}  \otimes_{b} \textbf{I}_{LN} \big)  E\Big( \boldsymbol{\mathcal{B}}^{T}(n) \otimes_{b} \boldsymbol{\mathcal{B}}^{T}(n) \Big) \hspace{0.2em} \boldsymbol{\sigma}
\end{split}
\end{equation}
\begin{equation}\label{eq4.3.11}
\begin{split}
\text{bvec}\Big\{ E \big[\boldsymbol{\mathcal{B}}^{T}(n)\hspace{0.1em} \boldsymbol{\Sigma} \hspace{0.1em} \boldsymbol{\mathcal{B}}(n) \big] \hspace{0.1em} \textbf{D} \hspace{0.1em} \boldsymbol{\eta} \overline{\boldsymbol{\mathcal{Q}}}_{M} \Big\}&= \big(  \overline{\boldsymbol{\mathcal{Q}}}_{M}^{T} \otimes_{b} \textbf{I}_{LN}  \big)  \big(  \boldsymbol{\eta} \otimes_{b} \textbf{I}_{LN}  \big) \big( \textbf{D}  \otimes_{b} \textbf{I}_{LN}\big) E\Big( \boldsymbol{\mathcal{B}}^{T}(n) \otimes_{b} \boldsymbol{\mathcal{B}}^{T}(n) \Big) \hspace{0.2em} \boldsymbol{\sigma}
\end{split}
\end{equation}
\begin{equation}\label{eq4.3.12}
\begin{split}
\text{bvec}\Big\{ \overline{\textbf{Z}}  \textbf{D}  E \big[\boldsymbol{\mathcal{B}}^{T}(n)\hspace{0.1em} \boldsymbol{\Sigma} \hspace{0.1em} \boldsymbol{\mathcal{B}}(n) \big] \Big\} &= \big(\textbf{I}_{LN} \otimes_{b} \overline{\textbf{Z}} \big) \big(\textbf{I}_{LN} \otimes_{b} \textbf{D}  \big) E\Big( \boldsymbol{\mathcal{B}}^{T}(n) \otimes_{b} \boldsymbol{\mathcal{B}}^{T}(n) \Big) \hspace{0.2em} \boldsymbol{\sigma}
\end{split}
\end{equation}
\begin{equation}\label{eq4.3.13}
\begin{split}
\text{bvec}\Big\{E\Big[ \textbf{Z}(n) \hspace{0.2em} \textbf{D}  \hspace{0.2em} \boldsymbol{\mathcal{B}}^{T}(n) \hspace{0.1em} \boldsymbol{\Sigma} \hspace{0.1em} \boldsymbol{\mathcal{B}}(n) \hspace{0.2em} \textbf{D}  \hspace{0.2em} \textbf{Z}(n)\Big] \Big\}
&= E\big(\textbf{Z}(n) \otimes_{b}  \textbf{Z}(n) \big)  \big(\textbf{D}  \otimes_{b} \textbf{D}  \big) E\Big( \boldsymbol{\mathcal{B}}^{T}(n) \otimes_{b} \boldsymbol{\mathcal{B}}^{T}(n) \Big) \hspace{0.2em} \boldsymbol{\sigma}
\end{split}
\end{equation}
\begin{equation}\label{eq4.3.14}
\begin{split}
\text{bvec}\Big\{ E\Big[ \textbf{Z}(n) \hspace{0.2em} \textbf{D}  \hspace{0.2em} \boldsymbol{\mathcal{B}}^{T}(n) \hspace{0.1em} \boldsymbol{\Sigma} \hspace{0.1em} \boldsymbol{\mathcal{B}}(n) \hspace{0.2em} \textbf{D}  \boldsymbol{\eta}\hspace{0.2em} \boldsymbol{\mathcal{Q}}_{M}(n)\Big] \Big \}
&= \big( \textbf{I}_{LN} \otimes_{b} \overline{\textbf{Z}}  \big) \big(\overline{\boldsymbol{\mathcal{Q}}}_{M}^{T}  \otimes_{b} \textbf{I}_{LN}  \big) \big(\boldsymbol{\eta} \otimes_{b} \textbf{I}_{LN}\big)  \big(\textbf{D}  \otimes_{b} \textbf{D} \big) \\ & \hspace{5em} E\Big( \boldsymbol{\mathcal{B}}^{T}(n) \otimes_{b} \boldsymbol{\mathcal{B}}^{T}(n) \Big) \hspace{0.2em} \boldsymbol{\sigma}
\end{split}
\end{equation}
\begin{equation}\label{eq4.3.15}
\begin{split}
\text{bvec}\Big\{ E\Big[\boldsymbol{\mathcal{Q}}_{M}^{T}(n) \hspace{0.2em} \boldsymbol{\eta} \hspace{0.2em} \textbf{D}  \boldsymbol{\mathcal{B}}^{T}(n)\hspace{0.1em} \boldsymbol{\Sigma} \hspace{0.1em} \boldsymbol{\mathcal{B}}(n) \Big] \Big  \}
&= \big( \textbf{I}_{LN} \otimes_{b} \overline{\boldsymbol{\mathcal{Q}}}_{M}^{T}\big) \big(\textbf{I}_{LN} \otimes_{b} \boldsymbol{\eta} \big) \big(\textbf{I}_{LN} \otimes_{b}  \textbf{D}  \big) E\Big( \boldsymbol{\mathcal{B}}^{T}(n) \otimes_{b} \boldsymbol{\mathcal{B}}^{T}(n) \Big) \hspace{0.2em} \boldsymbol{\sigma}
\end{split}
\end{equation}
\begin{equation}\label{eq4.3.16}
\begin{split}
\text{bvec}\Big\{ E\Big[\boldsymbol{\mathcal{Q}}_{M}^{T}(n) \hspace{0.2em} \boldsymbol{\eta} \hspace{0.2em} \textbf{D}  \boldsymbol{\mathcal{B}}^{T}(n)\hspace{0.1em} \boldsymbol{\Sigma} \hspace{0.1em} \boldsymbol{\mathcal{B}}(n) \textbf{D}  \textbf{Z}(n)\Big] \Big \}
&= \big( \overline{\textbf{Z}} \otimes_{b} \textbf{I}_{LN} \big) \big( \textbf{I}_{LN} \otimes_{b} \overline{\boldsymbol{\mathcal{Q}}}_{M}^{T}\big) \big(\textbf{I}_{LN} \otimes_{b} \boldsymbol{\eta} \big) \big(\textbf{D}  \otimes_{b} \textbf{D}  \big)  E\Big( \boldsymbol{\mathcal{B}}^{T}(n) \otimes_{b} \boldsymbol{\mathcal{B}}^{T}(n) \Big) \hspace{0.2em} \boldsymbol{\sigma}
\end{split}
\end{equation}
\begin{equation}\label{eq4.3.17}
\begin{split}
\text{bvec}\Big\{ E\Big[\boldsymbol{\mathcal{Q}}_{M}^{T}(n) \hspace{0.2em} \boldsymbol{\eta} \hspace{0.2em} \textbf{D}  \boldsymbol{\mathcal{B}}^{T}(n)\hspace{0.1em} \boldsymbol{\Sigma} \hspace{0.1em} \boldsymbol{\mathcal{B}}(n) \textbf{D}  \boldsymbol{\eta}  \hspace{0.2em} \boldsymbol{\mathcal{Q}}_{M}(n)\Big] \Big \}
&= E\big(  \boldsymbol{\mathcal{Q}}_{M}^{T}(n) \otimes_{b} \boldsymbol{\mathcal{Q}}_{M}^{T}(n)  \big) \big(\boldsymbol{\eta}^{T} \otimes_{b} \boldsymbol{\eta}^{T} \big)  \big(\textbf{D}  \otimes_{b} \textbf{D}  \big)  \\
& \hspace{5em} E\Big( \boldsymbol{\mathcal{B}}^{T}(n) \otimes_{b} \boldsymbol{\mathcal{B}}^{T}(n) \Big) \hspace{0.2em} \boldsymbol{\sigma}
\end{split}
\end{equation}
Therefore, a linear relation between the corresponding vectors $\{\boldsymbol{\sigma},\boldsymbol{\sigma}^{'}\}$is formulated by
\begin{equation}\label{eq4.3.18}
\begin{split}
\boldsymbol{\sigma}^{'}= \textbf{F} \boldsymbol{\sigma}
\end{split}
\end{equation}
where $\textbf{F}$ is an $L^{2}N^{2} \times L^{2}N^{2}$ matrix and given by
\begin{equation}\label{eq4.3.19}
\begin{split}
\textbf{F}&=\left[\begin{array}{l}  \textbf{I}_{LN} - \big( \overline{\textbf{Z}} \otimes_{b} \textbf{I}_{LN} \big)  \big(  \textbf{D}  \otimes_{b} \textbf{I}_{LN} \big) + \big(  \overline{\boldsymbol{\mathcal{Q}}}_{M}^{T} \otimes_{b} \textbf{I}_{LN}  \big)  \big(  \boldsymbol{\eta} \otimes_{b} \textbf{I}_{LN}  \big) \big( \textbf{D}  \otimes_{b} \textbf{I}_{LN}\big)  \\
-  \big(\textbf{I}_{LN} \otimes_{b} \overline{\textbf{Z}} \big) \big(\textbf{I}_{LN} \otimes_{b}  \textbf{D}  \big)
 +  E\big(\textbf{Z}(n) \otimes_{b}  \textbf{Z}(n) \big)  \big(\textbf{D}  \otimes_{b} \textbf{D} \big) \\
 - \big( \textbf{I}_{LN} \otimes_{b} \overline{\textbf{Z}}  \big) \big(\overline{\boldsymbol{\mathcal{Q}}}_{M}^{T}  \otimes_{b} \textbf{I}_{LN}  \big) \big(\boldsymbol{\eta} \otimes_{b} \textbf{I}_{LN}\big)  \big(\textbf{D}  \otimes_{b} \textbf{D} \big)  \\
  + \big( \textbf{I}_{LN} \otimes_{b} \overline{\boldsymbol{\mathcal{Q}}}_{M}^{T}\big) \big(\textbf{I}_{LN} \otimes_{b} \boldsymbol{\eta} \big) \big(\textbf{I}_{LN} \otimes_{b} \textbf{D}  \big) \\
 -   \big( \overline{\textbf{Z}} \otimes_{b} \textbf{I}_{LN} \big) \big( \textbf{I}_{LN} \otimes_{b} \overline{\boldsymbol{\mathcal{Q}}}_{M}^{T}\big) \big(\textbf{I}_{LN} \otimes_{b} \boldsymbol{\eta} \big) \big(\textbf{D}  \otimes_{b} \textbf{D}  \big) \\
 +  E\big(  \boldsymbol{\mathcal{Q}}_{M}^{T}(n) \otimes_{b} \boldsymbol{\mathcal{Q}}_{M}^{T}(n)  \big) \big(\boldsymbol{\eta}^{T} \otimes_{b} \boldsymbol{\eta}^{T} \big)  \big(\textbf{D}  \otimes_{b} \textbf{D}  \big)   \end{array} \right] E\Big( \boldsymbol{\mathcal{B}}^{T}(n) \otimes_{b} \boldsymbol{\mathcal{B}}^{T}(n) \Big)
\end{split}
\end{equation}
The evaluation of $\textbf{F}$ involves evaluating mainly two quantities that are $\boldsymbol{\Phi} = E\Big( \boldsymbol{\mathcal{B}}^{T}(n) \otimes_{b} \boldsymbol{\mathcal{B}}^{T}(n) \Big)$  and \\ $\boldsymbol{\Upsilon} = E\Big( \boldsymbol{\mathcal{Q}}_{M}^{T}(n) \otimes_{b} \boldsymbol{\mathcal{Q}}_{M}^{T}(n) \Big)$. First, $\boldsymbol{\Phi}$ is evaluated as follows:
\begin{equation}
\begin{split}
\boldsymbol{\Phi}&=E\Big( \boldsymbol{\mathcal{B}}^{T}(n) \otimes_{b} \boldsymbol{\mathcal{B}}^{T}(n) \Big)\\
&=E \bigg[ \Big( \boldsymbol{\mathcal{A}}  \odot  \boldsymbol{\mathcal{S}}^{T}(n) +  \Big( \textbf{I}_{LN} -  \boldsymbol{\mathcal{A}}^{T}    \boldsymbol{\mathcal{S}}^{T}(n) \Big) \odot \textbf{I}_{LN} \Big) \otimes_{b}  \Big( \boldsymbol{\mathcal{A}}  \odot  \boldsymbol{\mathcal{S}}^{T}(n) +  \Big( \textbf{I}_{LN} -  \boldsymbol{\mathcal{A}}^{T}    \boldsymbol{\mathcal{S}}^{T}(n) \Big) \odot \textbf{I}_{LN} \Big)   \bigg]\\
&= E \Big[ \Big(\boldsymbol{\mathcal{A}}  \odot  \boldsymbol{\mathcal{S}}^{T}(n) \Big) \otimes_{b} \Big( \boldsymbol{\mathcal{A}}  \odot  \boldsymbol{\mathcal{S}}^{T}(n) \Big) \Big] + E \Big[ \textbf{I}_{LN} \otimes_{b}  \Big( \boldsymbol{\mathcal{A}}  \odot  \boldsymbol{\mathcal{S}}^{T}(n) \Big)  \Big] + E \Big[ \Big( \boldsymbol{\mathcal{A}}  \odot  \boldsymbol{\mathcal{S}}^{T}(n) \Big) \otimes_{b} \textbf{I}_{LN}  \Big]  \\
& \hspace{1em} - E \Big[ \Big( \boldsymbol{\mathcal{A}}  \odot  \boldsymbol{\mathcal{S}}^{T}(n) \Big) \otimes_{b} \Big( \big(   \boldsymbol{\mathcal{A}}^{T}  \boldsymbol{\mathcal{S}}^{T}(n) \big) \odot \textbf{I}_{LN} \Big)  \Big]  - E \Big[  \Big( \big( \boldsymbol{\mathcal{A}}^{T}  \boldsymbol{\mathcal{S}}^{T}(n) \big) \odot \textbf{I}_{LN} \Big)  \otimes_{b} \Big( \boldsymbol{\mathcal{A}}  \odot  \boldsymbol{\mathcal{S}}^{T}(n) \Big) \Big]\\
& \hspace{1em} + \textbf{I}_{L^{2}N^{2}} - E \Big[ \textbf{I}_{LN} \otimes_{b} \Big( \big( \boldsymbol{\mathcal{A}}^{T}  \boldsymbol{\mathcal{S}}^{T}(n) \big) \odot \textbf{I}_{LN} \Big)  \Big]  - E \Big[  \Big( \big(\boldsymbol{\mathcal{A}}^{T}  \boldsymbol{\mathcal{S}}^{T}(n) \big) \odot \textbf{I}_{LN} \Big)  \otimes_{b} \textbf{I}_{LN} \Big]\\
& \hspace{1em} + E \Big[  \Big( \big( \boldsymbol{\mathcal{A}}^{T}  \boldsymbol{\mathcal{S}}^{T}(n) \big) \odot \textbf{I}_{LN} \Big)  \otimes_{b} \Big( \big( \boldsymbol{\mathcal{A}}^{T}  \boldsymbol{\mathcal{S}}^{T}(n) \big) \odot \textbf{I}_{LN} \Big) \Big]
\end{split}
\end{equation}
By defining $\boldsymbol{\Omega}_{S}= E\big(\boldsymbol{\mathcal{S}}^{T}(n) \otimes_{b} \boldsymbol{\mathcal{S}}^{T}(n) \big)$
we can evaluate the terms as follows:
\begin{equation}
\begin{split}
E \Big[ \Big(\boldsymbol{\mathcal{A}}  \odot  \boldsymbol{\mathcal{S}}^{T}(n) \Big) \otimes_{b} \Big( \boldsymbol{\mathcal{A}}  \odot  \boldsymbol{\mathcal{S}}^{T}(n) \Big) \Big] &=  \Big( \boldsymbol{\mathcal{A}}  \otimes_{b}   \boldsymbol{\mathcal{A}}  \Big) \odot \boldsymbol{\Omega}_{S}
\end{split}
\end{equation}
\begin{equation}
\begin{split}
E \Big[ \textbf{I}_{LN} \otimes_{b}  \Big( \boldsymbol{\mathcal{A}}\odot  \boldsymbol{\mathcal{S}}^{T}(n) \Big)  \Big] = \textbf{I}_{LN} \otimes_{b}  \Big( \boldsymbol{\mathcal{A}}\odot E\big( \boldsymbol{\mathcal{S}}^{T}(n) \big) \Big) = p \Big(\textbf{I}_{LN}  \otimes_{b} \boldsymbol{\mathcal{A}} \Big)
\end{split}
\end{equation}
\begin{equation}
\begin{split}
E \Big[ \Big( \boldsymbol{\mathcal{A}} \odot  \boldsymbol{\mathcal{S}}^{T}(n) \Big) \otimes_{b} \textbf{I}_{LN}  \Big]  = \Big( \boldsymbol{\mathcal{A}}  \odot E\big( \boldsymbol{\mathcal{S}}^{T}(n) \big) \Big)  \textbf{I}_{LN} \otimes_{b} = p \Big( \boldsymbol{\mathcal{A}}  \otimes_{b}  \textbf{I}_{LN}\Big)
\end{split}
\end{equation}
\begin{equation}
\begin{split}
E \Big[ \Big( \boldsymbol{\mathcal{A}} \odot  \boldsymbol{\mathcal{S}}^{T}(n) \Big) \otimes_{b} \Big( \big(   \boldsymbol{\mathcal{A}}^{T} \boldsymbol{\mathcal{S}}^{T}(n) \big) \odot \textbf{I}_{LN} \Big)  \Big] &= E \Big[ \Big( \boldsymbol{\mathcal{A}} \odot  \boldsymbol{\mathcal{S}}^{T}(n) \Big) \otimes_{b} \Big( \big( \textbf{I}_{LN}    \odot \boldsymbol{\mathcal{A}}^{T} \boldsymbol{\mathcal{S}}^{T}(n) \big) \Big)  \Big]\\
&= \Big( \boldsymbol{\mathcal{A}} \otimes_{b} \textbf{I}_{LN} \Big) \odot \Big( \textbf{I}_{LN} \otimes_{b} \boldsymbol{\mathcal{A}}^{T} \Big) \boldsymbol{\Omega}_{S}
\end{split}
\end{equation}
\begin{equation}
\begin{split}
E \Big[ \Big( \big(   \boldsymbol{\mathcal{A}}^{T} \boldsymbol{\mathcal{S}}^{T}(n) \big) \odot \textbf{I}_{LN} \Big) \otimes_{b} \Big( \boldsymbol{\mathcal{A}} \odot  \boldsymbol{\mathcal{S}}^{T}(n) \Big)   \Big] &= E \Big[ \Big( \big( \textbf{I}_{LN}    \odot \boldsymbol{\mathcal{A}}^{T} \boldsymbol{\mathcal{S}}^{T}(n) \big) \Big) \otimes_{b} \Big( \boldsymbol{\mathcal{A}} \odot  \boldsymbol{\mathcal{S}}^{T}(n) \Big) \Big]\\
&= \Big( \textbf{I}_{LN}  \otimes_{b}  \boldsymbol{\mathcal{A}} \Big) \odot \Big( \boldsymbol{\mathcal{A}}^{T}  \otimes \textbf{I}_{LN}   \Big) \boldsymbol{\Omega}_{S}
\end{split}
\end{equation}
\begin{equation}
\begin{split}
E \Big[ \textbf{I}_{LN} \otimes_{b} \Big( \big( \boldsymbol{\mathcal{A}}^{T}  \boldsymbol{\mathcal{S}}^{T}(n) \big) \odot \textbf{I}_{LN} \Big)  \Big] &= \textbf{I}_{LN} \otimes_{b} \Big( \big(  \boldsymbol{\mathcal{A}}  \hspace{0.2em} \overline{\boldsymbol{\mathcal{S}}} \big) \odot \textbf{I}_{LN} \Big)\\ &= \textbf{I}_{LN} \otimes_{b} p \Big( \big(  \textbf{A}  \otimes \textbf{I}_{L} \big) \hspace{0.2em} \big(\textbf{J}_{N} \otimes \textbf{I}_{L}\big) \odot \textbf{I}_{LN} \Big)\\
&=\textbf{I}_{LN} \otimes_{b} p \Big( \big(\textbf{J}_{N} \otimes \textbf{I}_{L}\big) \odot \textbf{I}_{LN} \Big)\\
&= p \hspace{0.2em} \textbf{I}_{L^{2}N^{2}}
\end{split}
\end{equation}
\begin{equation}
\begin{split}
E \Big[ \Big( \big( \boldsymbol{\mathcal{A}}^{T}  \boldsymbol{\mathcal{S}}^{T}(n) \big) \odot \textbf{I}_{LN} \Big)   \otimes_{b} \textbf{I}_{LN}\Big]
&= p \hspace{0.2em} \textbf{I}_{L^{2}N^{2}}
\end{split}
\end{equation}

\begin{equation}
\begin{split}
E \Big[  \Big( \big( \boldsymbol{\mathcal{A}}^{T}  \boldsymbol{\mathcal{S}}^{T}(n) \big) \odot \textbf{I}_{LN} \Big)  \otimes_{b} \Big( \big( \boldsymbol{\mathcal{A}}^{T}  \boldsymbol{\mathcal{S}}^{T}(n) \big) \odot \textbf{I}_{LN} \Big) \Big] & = E \Big[  \Big( \big( \boldsymbol{\mathcal{A}}^{T}  \boldsymbol{\mathcal{S}}^{T}(n) \big) \otimes_{b} \big( \boldsymbol{\mathcal{A}} (n) \boldsymbol{\mathcal{S}}^{T}(n) \big)  \Big)  \odot \textbf{I}_{L^{2}N^{2}} \Big]\\
&=  \Big( \boldsymbol{\mathcal{A}}^{T}   \otimes_{b}  \boldsymbol{\mathcal{A}}^{T}    \Big) \boldsymbol{\Omega}_{S}  \odot \textbf{I}_{L^{2}N^{2}} \\
\end{split}
\end{equation}
Therefore using these results we have
\begin{equation}
\begin{split}
\boldsymbol{\Phi}&= \Big( \boldsymbol{\mathcal{A}}  \otimes_{b}   \boldsymbol{\mathcal{A}}  \Big) \odot \boldsymbol{\Omega}_{S} + p \Big(\textbf{I}_{LN}  \otimes_{b}  \boldsymbol{\mathcal{A}}  \Big) + p \Big(  \boldsymbol{\mathcal{A}}   \otimes_{b}  \textbf{I}_{LN}\Big) - \Big( \boldsymbol{\mathcal{A}}  \otimes_{b} \textbf{I}_{LN} \Big) \odot \Big( \textbf{I}_{LN} \otimes_{b} \boldsymbol{\mathcal{A}}^{T}  \Big) \boldsymbol{\Omega}_{S} \\
&-\Big( \textbf{I}_{LN}  \otimes_{b}  \boldsymbol{\mathcal{A}}  \Big) \odot \Big( \boldsymbol{\mathcal{A}}^{T}  \otimes \textbf{I}_{LN}   \Big) \boldsymbol{\Omega}_{S} + (1- 2p) \textbf{I}_{L^{2}N^{2}} + \Big( \boldsymbol{\mathcal{A}}^{T}   \otimes_{b}  \boldsymbol{\mathcal{A}}^{T}    \Big) \boldsymbol{\Omega}_{S}  \odot \textbf{I}_{L^{2}N^{2}}
\end{split}
\end{equation}
In the same way we can also evaluate the quantity $\boldsymbol{\Upsilon}=E\Big( \boldsymbol{\mathcal{Q}}^{T}(n) \otimes_{b} \boldsymbol{\mathcal{Q}}^{T}(n) \Big)$ as follows:
\begin{equation}
\begin{split}
\boldsymbol{\Upsilon}&=E\Big( \boldsymbol{\mathcal{Q}}_{M}^{T}(n) \otimes_{b} \boldsymbol{\mathcal{Q}}_{M}^{T}(n) \Big)\\
&=E \bigg[ \Big( \boldsymbol{\mathcal{P}}^{T}  \odot  \boldsymbol{\mathcal{S}}^{T}(n) -  \Big(  \boldsymbol{\mathcal{P}} \hspace{0.5em}   \boldsymbol{\mathcal{S}}^{T}(n) \Big) \odot \textbf{I}_{LN} \Big) \otimes_{b}  \Big( \boldsymbol{\mathcal{P}}^{T}  \odot  \boldsymbol{\mathcal{S}}^{T}(n) - \Big( \boldsymbol{\mathcal{P}}    \boldsymbol{\mathcal{S}}^{T}(n) \Big) \odot \textbf{I}_{LN} \Big)   \bigg]\\
&= \Big( \boldsymbol{\mathcal{P}}  \otimes_{b}   \boldsymbol{\mathcal{P}}  \Big) \odot \boldsymbol{\Omega}_{S}  -\Big( \boldsymbol{\mathcal{P}}^{T}(n) \otimes_{b} \textbf{I}_{LN} \Big) \odot \Big( \textbf{I}_{LN} \otimes_{b} \boldsymbol{\mathcal{P}}(n) \Big) \boldsymbol{\Omega}_{s} \\
&\hspace{2em} - \Big( \textbf{I}_{LN}  \otimes_{b}  \boldsymbol{\mathcal{P}}^{T}(n) \Big) \odot \Big( \boldsymbol{\mathcal{P}}(n) \otimes \textbf{I}_{LN}   \Big) \boldsymbol{\Omega}_{s} + \bigg(  \Big(\textbf{C}_{p} + \big( \overline{\textbf{P}} \otimes \overline{\textbf{P}} \big)\Big) \otimes  \textbf{I}_{L^{2}}   \bigg) \boldsymbol{\Omega}_{s}  \odot \textbf{I}_{L^{2}N^{2}}
\end{split}
\end{equation}
Finally, the quantity $\boldsymbol{\Omega}_{S}=\boldsymbol{\mathcal{S}}^{T}(n)  \otimes_{b}  \boldsymbol{\mathcal{S}}^{T}(n)$ can be calculated as follows:
\begin{equation}
\begin{split}
\boldsymbol{\Omega}_{S}=\boldsymbol{\mathcal{S}}^{T}(n)  \otimes_{b}  \boldsymbol{\mathcal{S}}^{T}(n) = \textbf{1}^{T} \otimes E \left[ \begin{array}{l} \textbf{S}_{1}(n) \otimes_{b} \boldsymbol{\mathcal{S}}^{T}(n)    \\ \hspace{2em} \vdots \\ \textbf{S}_{N}(n) \otimes_{b} \boldsymbol{\mathcal{S}}^{T}(n) \end{array}\right]
\end{split}
\end{equation}
where
\begin{equation}
\begin{split}
 E \left[ \textbf{S}_{i}(n) \otimes_{b} \boldsymbol{\mathcal{S}}^{T}(n)   \right] = \text{blockdiag}\Big( s_{1, i}(n) \boldsymbol{\mathcal{S}}^{T}(n), s_{2, i}(n) \boldsymbol{\mathcal{S}}^{T}(n), \cdots, s_{L, i}(n) \boldsymbol{\mathcal{S}}^{T}(n)  \Big)  \hspace{2em} \text{for} \hspace{1em} i=1,2, \cdots, N.
\end{split}
\end{equation}
and
\begin{equation}
\begin{split}
E[s_{r, i}(n) \boldsymbol{\mathcal{S}}^{T}(n)] = \textbf{1}^{T} \otimes E \left[ \begin{array}{l} s_{r, i}(n) \textbf{S}_{1}(n)     \\ \hspace{2em} \vdots \\ s_{r, i}(n) \textbf{S}_{N}(n)  \end{array}\right],   \hspace{2em} \text{for} \hspace{1em} r=1,2, \cdots, L.
\end{split}
\end{equation}
with
\begin{equation}
\begin{split}
 E \left[ s_{r, i}(n) \textbf{S}_{j}(n)  \right] = \text{diag}\Big( E[ s_{r, i}(n) s_{1, j}(n) ], E[ s_{r, i}(n) s_{2, j}(n) ], \cdots, E[ s_{r, i}(n) s_{L, j}(n) ]  \Big)  \hspace{2em} \text{for} \hspace{1em} j=1,2, \cdots, N.
\end{split}
\end{equation}
The probability that at a given time two different entries of a single node are transmitted is $ (\frac{M}{L}) (\frac{M-1}{L-1})$. Moreover, in the uncoordinate partial-diffusion scheme, at any given time instant, the probability that at a given time two different entries of two different nodes are transmitted is $p^{2}$. Therefore from $\cite{23}$, for uncoordinated partial diffusion scheme, we have
\begin{equation}
\begin{split}
E\Big[s_{r, i}(n) \hspace{0.2em}  s_{s, j}(n) \Big]=\begin{cases}
p    \hspace{2em} \text{if} \hspace{2em} i=j \hspace{2em} \text{and} \hspace{2em}  r=s  \\
p (\frac{M-1}{L-1})  \hspace{2em} \text{if} \hspace{2em} i=j \hspace{2em} \text{and} \hspace{2em}  r \neq s \\
p^{2}  \hspace{2em} \text{if} \hspace{2em} i\neq j
\end{cases}
\end{split}
\end{equation}
and alternatively, for the coordinated partial diffusion scheme, we have
\begin{equation}
\begin{split}
E\Big[s_{r, i}(n) \hspace{0.2em}  s_{s, j}(n) \Big]=\begin{cases}
p    \hspace{2em} \text{if} \hspace{2em} \hspace{2em}  r=s  \\
p (\frac{M-1}{L-1})  \hspace{2em} \text{if} \hspace{2em}  r \neq s \\
\end{cases}
\end{split}
\end{equation}
On the other hand, for periodic partial diffusion scheme, we have
\begin{equation}
\begin{split}
E \left[ s_{r, i}(n) \textbf{S}_{j}(n)  \right] &= p \hspace{0.2em} \textbf{I}_{L}\\
E[s_{r, i}(n) \boldsymbol{\mathcal{S}}^{T}(n)] &= p \hspace{0.2em} \textbf{J}_{N} \otimes \textbf{I}_{L}\\
E \left[ \textbf{S}_{i}(n) \otimes_{b} \boldsymbol{\mathcal{S}}^{T}(n)   \right] &=  p \hspace{0.2em} \textbf{I}_{L} \otimes  \textbf{J}_{N} \otimes \textbf{I}_{L}\\
E \left[ \boldsymbol{\mathcal{S}}^{T}(n)  \otimes_{b} \boldsymbol{\mathcal{S}}^{T}(n)   \right] &=  p \hspace{0.2em} \textbf{J}_{N} \otimes \textbf{I}_{L} \otimes_{b}  \textbf{J}_{N} \otimes \textbf{I}_{L}\\
\end{split}
\end{equation}
There fore, using the above results for periodic partial diffusion scheme we can write
\begin{equation}
\begin{split}
\boldsymbol{\Phi}&= (1- p) \textbf{I}_{L^{2}N^{2}} + p \hspace{0.2em}   \big( \boldsymbol{\mathcal{A}} \otimes_{b}  \boldsymbol{\mathcal{A}} \big) \\
\boldsymbol{\Upsilon}&=  p \hspace{0.2em} \Big( \textbf{I}_{L^{2}N^{2}} +  \hspace{0.2em}  \big( \boldsymbol{\mathcal{P}} \otimes_{b}  \boldsymbol{\mathcal{P}} \big) -  \hspace{0.2em} \big( \textbf{I}_{LN} \otimes_{b}  \boldsymbol{\mathcal{P}} \big)-  \hspace{0.2em}   \big( \boldsymbol{\mathcal{P}} \otimes_{b} \textbf{I}_{LN} \big)  \Big)= p \hspace{0.2em} \big(\boldsymbol{\mathcal{Q}}^{T} \otimes_{b} \boldsymbol{\mathcal{Q}}^{T} \big)
 \end{split}
\end{equation}
As proved in $\cite{11}$ we can prove that the sum of each row of $ \boldsymbol{\Phi}$ is equal to $1$. Using these results $F$ can be rewritten as,
\begin{equation}\label{eq4.3.19}
\begin{split}
\textbf{F}&=\left[\begin{array}{l}  \textbf{I}_{LN} - \big( \overline{\textbf{Z}} \otimes_{b} \textbf{I}_{LN} \big)  \big(  \textbf{D}  \otimes_{b} \textbf{I}_{LN} \big) \\
+ p \hspace{0.2em} \big(   \boldsymbol{\mathcal{Q}}^{T} \otimes_{b} \textbf{I}_{LN}  \big)  \big(  \boldsymbol{\eta} \otimes_{b} \textbf{I}_{LN}  \big) \big( \textbf{D}  \otimes_{b} \textbf{I}_{LN}\big) \\
 -  \big(\textbf{I}_{LN} \otimes_{b} \overline{\textbf{Z}} \big) \big(\textbf{I}_{LN} \otimes_{b}  \textbf{D}  \big)  +  E\big(\textbf{Z}(n) \otimes_{b}  \textbf{Z}(n) \big)  \big(\textbf{D}  \otimes_{b} \textbf{D} \big) \\
 - p \hspace{0.2em} \big( \textbf{I}_{LN} \otimes_{b} \overline{\textbf{Z}}  \big) \big( \boldsymbol{\mathcal{Q}}^{T}  \otimes_{b} \textbf{I}_{LN}  \big) \big(\boldsymbol{\eta} \otimes_{b} \textbf{I}_{LN}\big)  \big(\textbf{D}  \otimes_{b} \textbf{D} \big) \\
  +  p \hspace{0.2em} \big( \textbf{I}_{LN} \otimes_{b}  \boldsymbol{\mathcal{Q}}^{T}\big) \big(\textbf{I}_{LN} \otimes_{b} \boldsymbol{\eta} \big) \big(\textbf{I}_{LN} \otimes_{b} \textbf{D}  \big) \\
 -  p \hspace{0.2em} \big( \overline{\textbf{Z}} \otimes_{b} \textbf{I}_{LN} \big) \big( \textbf{I}_{LN} \otimes_{b} \boldsymbol{\mathcal{Q}}^{T}\big) \big(\textbf{I}_{LN} \otimes_{b} \boldsymbol{\eta} \big) \big(\textbf{D}  \otimes_{b} \textbf{D}  \big) \\
 + p \hspace{0.2em} E\big(  \boldsymbol{\mathcal{Q}}^{T} \otimes_{b} \boldsymbol{\mathcal{Q}}^{T} \big) \big(\boldsymbol{\eta}^{T} \otimes_{b} \boldsymbol{\eta}^{T} \big)  \big(\textbf{D}  \otimes_{b} \textbf{D}  \big)   \end{array} \right] \bigg( (1- p) \textbf{I}_{L^{2}N^{2}} + p \hspace{0.2em}   \big( \boldsymbol{\mathcal{A}} \otimes_{b}  \boldsymbol{\mathcal{A}} \big) \bigg)
\end{split}
\end{equation}
\par
Let $\Lambda_{v} = E[\textbf{\emph{v}}(n)\textbf{\emph{v}}^{T}(n)]$ denote a $NP \times NP$  diagonal matrix, whose entries
are the noise variances $\sigma^{2}_{v,k}$ for $k = 1,2, \cdots ,N$ and given by
\begin{equation}\label{eq4.3.20}
\begin{split}
\Lambda_{v}=\text{diag}\{\sigma^{2}_{v, 1}\textbf{I}_{p}, \sigma^{2}_{v, 2}\textbf{I}_{p}, \cdots, \sigma^{2}_{v, N}\textbf{I}_{p}\}
\end{split}
\end{equation}
Using the independence assumption of noise signals, the term $E\big[\textbf{\emph{v}}^{T}(n) \hspace{0.1em}\textbf{Y}^{\boldsymbol{\Sigma}}(n) \hspace{0.1em} \textbf{\emph{v}}(n) \big]$ can be written as
\begin{equation}\label{eq4.3.21}
\begin{split}
E\big[\textbf{\emph{v}}^{T}(n) \hspace{0.1em}\textbf{Y}^{\boldsymbol{\Sigma}}(n) \hspace{0.1em} \textbf{v}(n) \big]&= Tr\Big( E\Big[ \boldsymbol{\mathcal{B}}(n) \hspace{0.1em} \textbf{D}  \hspace{0.1em}E[\boldsymbol{\Phi}(n)]\hspace{0.1em} \textbf{D}^{T}  \hspace{0.1em}\boldsymbol{\mathcal{B}}^{T}(n) \Big] \boldsymbol{\Sigma} \Big)\\
&= \boldsymbol{\gamma}^{T} \hspace{0.1em} \boldsymbol{\sigma}
\end{split}
\end{equation}
where $\boldsymbol{\Phi}(n)= \textbf{U}^{T}(n)\big[\varepsilon \textbf{I} + \textbf{U}(n) \textbf{U}^{T}(n) \big]^{-1} \boldsymbol{\Lambda}_{v}(n) \big[\varepsilon \textbf{I} + \textbf{U}(n) \textbf{U}^{T}(n) \big]^{-1} \textbf{U}(n) $
and
\begin{equation}\label{eq4.3.22}
\begin{split}
\boldsymbol{\gamma}&= \text{vec}\Big\{ E\Big[ \boldsymbol{\mathcal{B}}(n) \hspace{0.1em} \textbf{D}  \hspace{0.1em}E[\boldsymbol{\Phi}(n)]\hspace{0.1em} \textbf{D}^{T}  \hspace{0.1em}\boldsymbol{\mathcal{B}}^{T}(n) \Big] \Big\}\\
&= E\big( \boldsymbol{\mathcal{B}}(n) \otimes \boldsymbol{\mathcal{B}}(n) \big)   \big(\textbf{D}  \otimes \textbf{D}  \big) \text{vec}\big\{ E[\textbf{W}^{T}(n) \boldsymbol{\Lambda}_{v} \textbf{W}(n) ]  \big\} \\
&= E\big( \boldsymbol{\mathcal{B}}(n) \otimes \boldsymbol{\mathcal{B}}(n) \big)   \big(\textbf{D}  \otimes \textbf{D}  \big)  E\Big(\textbf{W}^{T}(n) \otimes \textbf{W}^{T}(n) \Big) \hspace{0.2em} \boldsymbol{\gamma}_{v}
\end{split}
\end{equation}
with $\textbf{W}(n)=\big[\varepsilon \textbf{I} + \textbf{U}(n) \textbf{U}^{*}(n) \big]^{-1} \textbf{U}(n)$ and $\boldsymbol{\gamma}_{v}=\text{vec}\{\boldsymbol{\Lambda}_{v}\}$.
\par
Now, consider the term $E\|\textbf{r}(n) \|_{\boldsymbol{\Sigma}}^{2} $, that can be written
\begin{equation}
\begin{split}
E\|\textbf{r}(n) \|_{\boldsymbol{\Sigma}}^{2}&= \Big(\text{bvec}\Big\{E\Big[ \boldsymbol{\mathcal{B}}(n) \textbf{D} \hspace{0.2em} \boldsymbol{\eta} \hspace{0.2em} \boldsymbol{\mathcal{Q}}_{M}(n) \textbf{w}^{\star} \big(\textbf{w}^{\star}\big)^{T}  \boldsymbol{\mathcal{Q}}_{M}^{T}(n) \boldsymbol{\eta}  \hspace{0.2em} \textbf{D} \hspace{0.2em}  \boldsymbol{\mathcal{B}}^{T}(n)\Big]\Big\} \Big)^{T} \boldsymbol{\sigma}\\
&= \textbf{r}^{T}_{b} \boldsymbol{\sigma}
\end{split}
\end{equation}
where
\begin{equation}
\begin{split}
\textbf{r}_{b}&= E\Big( \boldsymbol{\mathcal{B}}(n) \otimes_{b} \boldsymbol{\mathcal{B}}(n) \Big) \big( \textbf{D} \otimes_{b} \textbf{D} \big) \big( \boldsymbol{\eta} \otimes_{b} \boldsymbol{\eta} \big) E\Big( \boldsymbol{\mathcal{Q}}_{M}(n) \otimes_{b} \boldsymbol{\mathcal{Q}}_{M}(n) \Big)  \text{bvec}\Big\{ \textbf{w}^{\star} \big(\textbf{w}^{\star}\big)^{T} \Big\}\\
&= p \hspace{0.2em}\bigg( (1- p) \textbf{I}_{L^{2}N^{2}} + p \hspace{0.2em}   \big( \boldsymbol{\mathcal{A}} \otimes_{b}  \boldsymbol{\mathcal{A}} \big) \bigg) \big( \textbf{D} \otimes_{b} \textbf{D} \big) \big( \boldsymbol{\eta} \otimes_{b} \boldsymbol{\eta} \big) \Big( \boldsymbol{\mathcal{Q}} \otimes_{b} \boldsymbol{\mathcal{Q}} \Big)  \text{bvec}\Big\{ \textbf{w}^{\star} \big(\textbf{w}^{\star}\big)^{T} \Big\}
\end{split}
\end{equation}
\par
Consider the term $E\Big[ \widetilde{\textbf{w}}^{T}(n) \hspace{0.2em}  \textbf{G}^{T}(n)  \hspace{0.2em}\boldsymbol{\Sigma} \hspace{0.3em}\textbf{r}(n) \Big]$ that can be simplified as follows:
\begin{equation}
\begin{split}
E\Big[ \widetilde{\textbf{w}}^{T}(n) \hspace{0.2em}  \textbf{G}^{T}(n)  \hspace{0.1em}\boldsymbol{\Sigma} \hspace{0.3em}\textbf{r}(n) \Big] &=
Tr\Big(E\Big[  \textbf{r}(n) \hspace{0.2em} \widetilde{\textbf{w}}^{T}(n) \hspace{0.3em}\textbf{G}^{T}(n) \Big] \hspace{0.1em}\boldsymbol{\Sigma} \Big)\\
&=\boldsymbol{\alpha}^{T}_{1}(n) \boldsymbol{\sigma}
\end{split}
\end{equation}
where
\begin{equation}
\begin{split}
\boldsymbol{\alpha}_{1}(n)&= E\Big( \boldsymbol{\mathcal{B}}(n) \otimes_{b} \boldsymbol{\mathcal{B}}(n) \Big) \left[\begin{array}{l}  \big( \textbf{I}_{LN} \otimes_{b}  \textbf{D}  \big)  \big( \textbf{I}_{LN} \otimes_{b} \boldsymbol{\eta} \big) \Big( \textbf{I}_{LN} \otimes_{b} \overline{\boldsymbol{\mathcal{Q}}}_{M} \Big) \\-  \big( \textbf{D}  \otimes_{b} \textbf{D}  \big)  \big( \textbf{I}_{LN} \otimes_{b} \boldsymbol{\eta} \big) \Big( \textbf{I}_{LN} \otimes_{b} \overline{\boldsymbol{\mathcal{Q}}}_{M} \Big) \big( \overline{\textbf{Z}} \otimes_{b} \textbf{I}_{LN} \big)  \\+  \big( \textbf{D}  \otimes_{b} \textbf{D}  \big) \big( \boldsymbol{\eta} \otimes_{b} \boldsymbol{\eta} \big) E\Big( \boldsymbol{\mathcal{Q}}_{M}(n) \otimes_{b} \boldsymbol{\mathcal{Q}}_{M}(n) \Big) \end{array}\right] \text{bvec}\Big\{\textbf{w}^{\star} E[\widetilde{\textbf{w}}^{T}(n) ] \}\\
&=p \hspace{0.2em}\bigg( (1- p) \textbf{I}_{L^{2}N^{2}} + p \hspace{0.2em}   \big( \boldsymbol{\mathcal{A}}^{T} \otimes_{b}  \boldsymbol{\mathcal{A}}^{T} \big) \bigg)  \left[\begin{array}{l}  \big( \textbf{I}_{LN} \otimes_{b}  \textbf{D}  \big)  \big( \textbf{I}_{LN} \otimes_{b} \boldsymbol{\eta} \big) \Big( \textbf{I}_{LN} \otimes_{b} \boldsymbol{\mathcal{Q}} \Big) \\-  \big( \textbf{D}  \otimes_{b} \textbf{D}  \big)  \big( \textbf{I}_{LN} \otimes_{b} \boldsymbol{\eta} \big) \Big( \textbf{I}_{LN} \otimes_{b} \boldsymbol{\mathcal{Q}} \Big) \big( \overline{\textbf{Z}} \otimes_{b} \textbf{I}_{LN} \big)  \\+  \big( \textbf{D}  \otimes_{b} \textbf{D}  \big) \big( \boldsymbol{\eta} \otimes_{b} \boldsymbol{\eta} \big) \Big( \boldsymbol{\mathcal{Q}} \otimes_{b} \boldsymbol{\mathcal{Q}} \Big) \end{array}\right] \text{bvec}\Big\{\textbf{w}^{\star} E[\widetilde{\textbf{w}}^{T}(n) ] \}\\
\end{split}
\end{equation}
\par
Same way we can write $E\Big[ \textbf{r}^{T}(n) \hspace{0.3em}\boldsymbol{\Sigma}\hspace{0.1em} \textbf{G}(n)\hspace{0.1em} \widetilde{\textbf{w}}(n)\Big]$ as follows:
\begin{equation}
\begin{split}
E\Big[ \textbf{r}^{T}(n) \hspace{0.3em}\boldsymbol{\Sigma}\hspace{0.1em} \textbf{G}(n)\hspace{0.1em} \widetilde{\textbf{w}}(n)\Big] &=
Tr\Big(E\Big[  \textbf{G}(n)\hspace{0.1em} \widetilde{\textbf{w}}(n) \hspace{0.2em}\textbf{r}^{T}(n) \Big] \hspace{0.1em}\boldsymbol{\Sigma} \Big)\\
&=\boldsymbol{\alpha}^{T}_{2}(n)  \boldsymbol{\sigma}
\end{split}
\end{equation}
where
\begin{equation}
\begin{split}
\boldsymbol{\alpha}_{2}(n)&= E\Big( \boldsymbol{\mathcal{B}}(n) \otimes_{b} \boldsymbol{\mathcal{B}}(n) \Big) \left[\begin{array}{l}  \big( \textbf{D}  \otimes_{b} \textbf{I}_{LN} \big)  \big( \boldsymbol{\eta} \otimes_{b} \textbf{I}_{LN} \big) \Big(  \overline{\boldsymbol{\mathcal{Q}}}_{M} \otimes_{b} \textbf{I}_{LN} \Big) \\
-  \big( \textbf{D}  \otimes_{b} \textbf{D}  \big)  \big( \boldsymbol{\eta} \otimes_{b} \textbf{I}_{LN} \big) \Big(  \overline{\boldsymbol{\mathcal{Q}}}_{M} \otimes_{b} \textbf{I}_{LN} \Big)  \big( \textbf{I}_{LN} \otimes_{b} \overline{\textbf{Z}} \big)  \\+  \big( \textbf{D}  \otimes_{b} \textbf{D}  \big) \big( \boldsymbol{\eta} \otimes_{b} \boldsymbol{\eta} \big) E\Big( \boldsymbol{\mathcal{Q}}_{M}(n) \otimes_{b} \boldsymbol{\mathcal{Q}}_{M}(n) \Big) \end{array}\right] \text{bvec}\Big\{ E[\widetilde{\textbf{w}}(n)] \big(\textbf{w}^{\star} \big)^{T}  \}\\
&=p \hspace{0.2em}\bigg( (1- p) \textbf{I}_{L^{2}N^{2}} + p \hspace{0.2em}   \big( \boldsymbol{\mathcal{A}}^{T} \otimes_{b}  \boldsymbol{\mathcal{A}}^{T} \big) \bigg)\left[\begin{array}{l}  \big( \textbf{D}  \otimes_{b} \textbf{I}_{LN} \big)  \big( \boldsymbol{\eta} \otimes_{b} \textbf{I}_{LN} \big) \Big(  \boldsymbol{\mathcal{Q}} \otimes_{b} \textbf{I}_{LN} \Big) \\
-  \big( \textbf{D}  \otimes_{b} \textbf{D}  \big)  \big( \boldsymbol{\eta} \otimes_{b} \textbf{I}_{LN} \big) \Big(  \boldsymbol{\mathcal{Q}} \otimes_{b} \textbf{I}_{LN} \Big)  \big( \textbf{I}_{LN} \otimes_{b} \overline{\textbf{Z}} \big)  \\+  \big( \textbf{D}  \otimes_{b} \textbf{D}  \big) \big( \boldsymbol{\eta} \otimes_{b} \boldsymbol{\eta} \big) \Big( \boldsymbol{\mathcal{Q}} \otimes_{b} \boldsymbol{\mathcal{Q}} \Big) \end{array}\right] \text{bvec}\Big\{ E[\widetilde{\textbf{w}}(n)] \big(\textbf{w}^{\star} \big)^{T}  \}\\
\end{split}
\end{equation}
\par
Therefore, let us define the $\emph{\textbf{f}}\big(\textbf{r}, E[\widetilde{\textbf{w}}(n)], \boldsymbol{\sigma}\big)$ as the last three terms on the right hand side of the $\eqref{eq4.3.3}$, i.e,
\begin{equation}\label{eq4.3.23}
\begin{split}
\emph{\textbf{f}}\big(\textbf{r}, E[\widetilde{\textbf{w}}(n)], \boldsymbol{\sigma}\big)&= E\|\textbf{r}(n) \|_{\boldsymbol{\Sigma}}^{2} - E\Big[ \widetilde{\textbf{w}}^{T}(n) \hspace{0.2em} \textbf{G}^{T}(n) \hspace{0.1em} \boldsymbol{\Sigma} \hspace{0.3em}\textbf{r}(n) \Big]- E\Big[ \textbf{r}^{T}(n) \hspace{0.3em}\boldsymbol{\Sigma}\hspace{0.1em} \textbf{G}(n)\hspace{0.1em} \widetilde{\textbf{w}}(n) \Big] \\
&= \big( \textbf{r}^{T}_{b} - \boldsymbol{\alpha}^{T}_{1}(n) - \boldsymbol{\alpha}^{T}_{2}(n) \big) \boldsymbol{\sigma}
\end{split}
\end{equation}
Therefore, from the above results the mean-square behavior of the multi-task partial diffusion APA algorithm is summarized
as follows:
\begin{equation}\label{eq4.3.24}
\begin{split}
E\|\widetilde{\textbf{w}}(n+1)\|_{\boldsymbol{\sigma}}^{2}&= E\|\widetilde{\textbf{w}}(n)\|_{\textbf{F}\boldsymbol{\sigma}}^{2}
+  \boldsymbol{\gamma}^{T}  \hspace{0.1em} \boldsymbol{\sigma} + \emph{\textbf{f}}\big(\textbf{r}, E[\widetilde{\textbf{w}}(n)], \boldsymbol{\sigma}\big)
\end{split}
\end{equation}
The proposed multi-task partial diffusion strategy presented in $\eqref{eq3.11}$ is mean square stable if the matrix $\textbf{F}$ is stable. Iterating the recursion $\eqref{eq4.3.24}$ starting from $n=0$, we get
\begin{equation}\label{eq4.3.25}
\begin{split}
E\|\widetilde{\textbf{w}}(n+1)\|_{\boldsymbol{\sigma}}^{2}&= E\|\widetilde{\textbf{w}}(0)\|_{\textbf{F}^{n+1}\boldsymbol{\sigma}}^{2}
+ \boldsymbol{\gamma}^{T} \hspace{0.1em} \sum\limits_{i=0}^{n}\textbf{F}^{i}\boldsymbol{\sigma}+\sum\limits_{i=0}^{n}\emph{\textbf{f}}\big(\textbf{r}, E[\widetilde{\textbf{w}}(n-i)], \textbf{F}^{i} \boldsymbol{\sigma}\big)
\end{split}
\end{equation}
with initial condition $\widetilde{\textbf{w}}(0)=\textbf{w}^{\star} - \textbf{w}(0)$. If the matrix $\textbf{F}$ is stable then the first and second terms in the above equation converge to a finite value as $n \rightarrow \infty $. Now, let us consider the third term on the RHS of the $\eqref{eq4.3.25}$. We know that $E[\widetilde{\textbf{w}}(n)]$ is uniformly bounded because $\eqref{eq4.2.5}$ is a BIBO stable recursion with bounded driving term $ \hspace{0.1em} \overline{\boldsymbol{\mathcal{B}}} \hspace{0.1em}\textbf{D} \hspace{0.1em} \boldsymbol{\eta} \hspace{0.1em}\overline{\boldsymbol{\mathcal{Q}}}_{M} \hspace{0.1em} \textbf{w}^{\star}= p \hspace{0.1em} \overline{\boldsymbol{\mathcal{B}}} \hspace{0.1em}\textbf{D} \hspace{0.1em} \boldsymbol{\eta} \hspace{0.1em} \boldsymbol{\mathcal{Q}} \hspace{0.1em} \textbf{w}^{\star} $. Therefore, from $\eqref{eq4.3.23}$  $\emph{\textbf{f}}\big(\textbf{r}, E[\widetilde{\textbf{w}}(n-i)], \textbf{F}^{i} \boldsymbol{\sigma}\big)$ can be written as
\begin{equation}\label{eq4.3.26}
\begin{split}
\emph{\textbf{f}}\big(\textbf{r}, E[\widetilde{\textbf{w}}(n-i)], \textbf{F}^{i} \boldsymbol{\sigma}\big)&= \big( \textbf{r}^{T}_{b} - \boldsymbol{\alpha}^{T}_{1}(n-i) - \boldsymbol{\alpha}^{T}_{2}(n-i) \big) \textbf{F}^{i} \boldsymbol{\sigma}
\end{split}
\end{equation}
Provided that $\textbf{F}$ is stable and there exist a matrix norm, denoted by $\|\cdot\|_{p}$ such that $\|\textbf{F}\|_{p}=c_{p} < 1$. Applying this norm to $\emph{\textbf{f}}$ and using the matrix norms and triangular inequality, we can write $ \| \emph{\textbf{f}}\big(\textbf{r}, E[\widetilde{\textbf{w}}(n-i)], \textbf{F}^{i} \boldsymbol{\sigma}\big)  \| \leq \emph{v} \hspace{0.1em} c^{i}_{p}$, given $\emph{v}$ is a small positive constant. Therefore $E\|\widetilde{\textbf{w}}(n+1)\|_{\boldsymbol{\sigma}}^{2}$ converges to a bounded value as $n \rightarrow \infty$, and the algorithm is said to be mean square stable.
\par
By selecting $\boldsymbol{\Sigma}=\frac{1}{N} \textbf{I}_{LN}$ we can relate $E\|\widetilde{\textbf{w}}(n+1)\|_{\boldsymbol{\sigma}}^{2}$ and $E\|\widetilde{\textbf{w}}(n)\|_{\boldsymbol{\sigma}}^{2}$ as follows:
\begin{equation}\label{eq4.3.27}
\begin{split}
E\|\widetilde{\textbf{w}}(n+1)\|_{\boldsymbol{\sigma}}^{2}&=E\|\widetilde{\textbf{w}}(n)\|_{\boldsymbol{\sigma}}^{2} + \boldsymbol{\gamma}^{T} \textbf{F}^{n} \boldsymbol{\sigma} - E\|\widetilde{\textbf{w}}(0)\|_{\big(I_{(LN)^{2}} - \textbf{F} \hspace{0.1em}\big)\textbf{F}^{n} \boldsymbol{\sigma}}^{2} + \sum\limits_{i=0}^{n}\emph{\textbf{f}}\big(\textbf{r}, E[\widetilde{\textbf{w}}(n-i)], \textbf{F}^{i}\boldsymbol{\sigma}\big) \\
& \hspace{2em} - \sum\limits_{i=0}^{n-1}\emph{\textbf{f}}\big(\textbf{r}, E[\widetilde{\textbf{w}}(n-1-i)], \textbf{F}^{i}\boldsymbol{\sigma}\big)
\end{split}
\end{equation}
we can rewrite the last two terms in the above equation as,
\begin{equation}\label{eq4.3.28}
\begin{split}
&\sum\limits_{i=0}^{n}\emph{\textbf{f}}\big(\textbf{r}, E[\widetilde{\textbf{w}}(n-i)], \textbf{F}^{i}\boldsymbol{\sigma}\big)
 - \sum\limits_{i=0}^{n-1}\emph{\textbf{f}}\big(\textbf{r}, E[\widetilde{\textbf{w}}(n-1-i)], \textbf{F}^{i}\boldsymbol{\sigma}\big) =  \textbf{r}_{b}^{T} \textbf{F}^{n} \hspace{0.1em} \boldsymbol{\sigma} - \left[   \boldsymbol{\alpha}_{1}^{T}(n) + \boldsymbol{\alpha}_{2}^{T}(n)   + \Gamma(n)  \right] \hspace{0.1em}  \boldsymbol{\sigma}
\end{split}
\end{equation}
where
\begin{equation}\label{eq4.3.29}
\begin{split}
\boldsymbol{\Gamma} (n) =   \sum\limits_{i=1}^{n}  \boldsymbol{\alpha}_{1}^{T}(n-i) +  \boldsymbol{\alpha}_{2}^{T}(n-i) \textbf{F}^{i} \boldsymbol{\sigma} - \sum\limits_{i=0}^{n-1}  \boldsymbol{\alpha}_{1}^{T}(n-1-i) +  \boldsymbol{\alpha}_{2}^{T}(n-1-i) \textbf{F}^{i} \boldsymbol{\sigma}
\end{split}
\end{equation}
Therefore, the recursion presented in $\eqref{eq4.3.24}$ can be rewritten as,
\begin{equation}\label{eq4.3.30}
\begin{split}
E\|\widetilde{\textbf{w}}(n+1)\|_{\boldsymbol{\sigma}}^{2}&= E\|\widetilde{\textbf{w}}(n)\|_{\boldsymbol{\sigma}}^{2} + \boldsymbol{\gamma}^{T} \textbf{F}^{n} \boldsymbol{\sigma} - E\|\widetilde{\textbf{w}}(0)\|_{\big(I_{(LN)^{2}} - \textbf{F} \hspace{0.1em}\big)\textbf{F}^{n} \boldsymbol{\sigma}}^{2} +\textbf{r}_{b}^{T} \textbf{F}^{n} \hspace{0.1em} \boldsymbol{\sigma} - \left[   \boldsymbol{\alpha}_{1}^{T}(n) + \boldsymbol{\alpha}_{2}^{T}(n)   + \Gamma(n)  \right] \\
\boldsymbol{\Gamma} (n+1)&= \boldsymbol{\Gamma} (n) \textbf{F} + \Big[  \big[ \boldsymbol{\alpha}_{1}^{T}(n) + \boldsymbol{\alpha}_{2}^{T}(n)  \big]\hspace{0.2em} [\textbf{F}- \textbf{I}_{(LN)^{2}}]     \Big]
\end{split}
\end{equation}
with $\boldsymbol{\Gamma} (0)= 0_{1 \times (LN)^{2}}$.
\par
Steady-state MSD of the multi-task partial diffusion APA strategy over asynchronous network is given as follows
\begin{equation}\label{eq4.3.31}
\begin{split}
\lim\limits_{n \to \infty}E\|\widetilde{\textbf{w}}(n)\|_{E\big[I_{(LN)^{2}}-\textbf{F}\big] \boldsymbol{\sigma}}^{2}&= \boldsymbol{\gamma}^{T}  \boldsymbol{\sigma} + \emph{\textbf{f}}\big(\textbf{r}, E[\widetilde{\textbf{w}}(\infty)], \boldsymbol{\sigma} \big)
\end{split}
\end{equation}
\section{Simulation Results}
A network consists of $9$ nodes with the topology shown in Fig. 1 was considered for simulations. The nodes were grouped into $3$ clusters: $\mathcal{C}_{1} = \{1, 2, 3\}, \mathcal{C}_{2} = \{4, 5, 6\},$ and $\mathcal{C}_{3} = \{7, 8, 9\}$. To evaluate the performance of the proposed multi-task partial diffusion APA, randomly generated coefficient vectors of length $L=256$ taps are considered for simulations. Randomly generated coefficient vectors of the form $\textbf{w}^{*}_{\mathcal{C}_{k}}=\textbf{w}_{0}+ \delta_{\mathcal{C}_{k}} \textbf{w}_{\mathcal{C}_{k}}$ with $L=256$ taps length were chosen as $\delta_{\mathcal{C}_{1}}=0.025, \delta_{\mathcal{C}_{2}}=-0.025$ and $\delta_{\mathcal{C}_{3}}=0.015$. Regularization strength $\rho_{kl}$ was set to $\rho_{kl}= |\mathcal{N}_{k}\setminus\mathcal{C}(k)|^{-1}$ for $l \in \mathcal{N}_{k}\setminus\mathcal{C}(k)$, and $\rho_{kl}=0$ for any other $l$. This settings usually leads to asymmetrical regularization weights. The coefficient matrix $C$ was taken to be identity matrix and the combiner coefficients $a_{lk}$ were set according to Metropolis rule.\\
\begin{figure}[h]
\centering
\includegraphics [height=65mm,width=80mm]{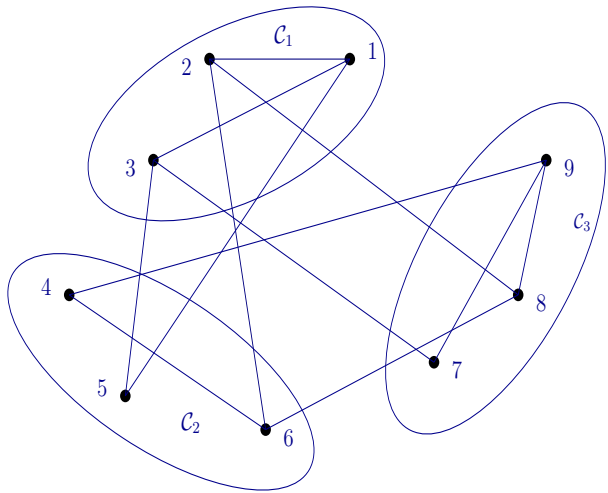}
\caption{Network Topology}
\label{the-label-for-cross-referencing}
\end{figure}
The input regressors $\textbf{u}_{k}(n)$ were taken from zero mean, Gaussian distribution with correlation statistics as shown in the Fig. 2 and the observation noises were i. i. d zero-mean Gaussian random variables, independent of any other signals with noise variances as shown in the Fig. 3. The multi-task partial diffusion APA algorithm was run with different values of $M$ i.e, the number of coefficients exchanged among nodes.
\begin{figure}[h]
\centering
\includegraphics [height=30mm,width=90mm]{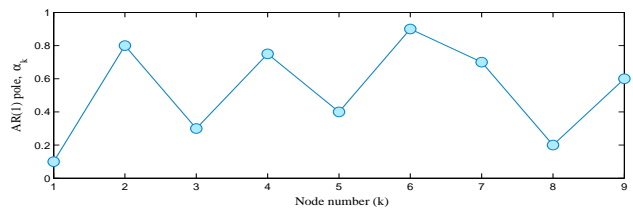}
\caption{Input statistics}
\label{Input statistics}
\end{figure}
\begin{figure}[h]
\centering
\includegraphics [height=30mm,width=90mm]{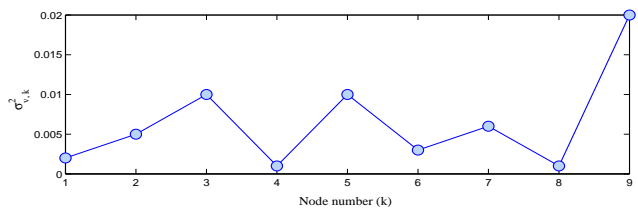}
\caption{Noise statistics}
\label{Noise statistics}
\end{figure}
\par
Projection order was taken to be $8$ and the initial taps were chosen to be zero. Normalized MSD was taken as the performance parametric to compare the diffusion strategies. The regularization parameter $\eta =0.0018$ was maintained same value. Simulation results were obtained by averaging $50$ Monte-Carlo runs.
\begin{figure}[h]
\centering
\includegraphics [height=85mm,width=100mm]{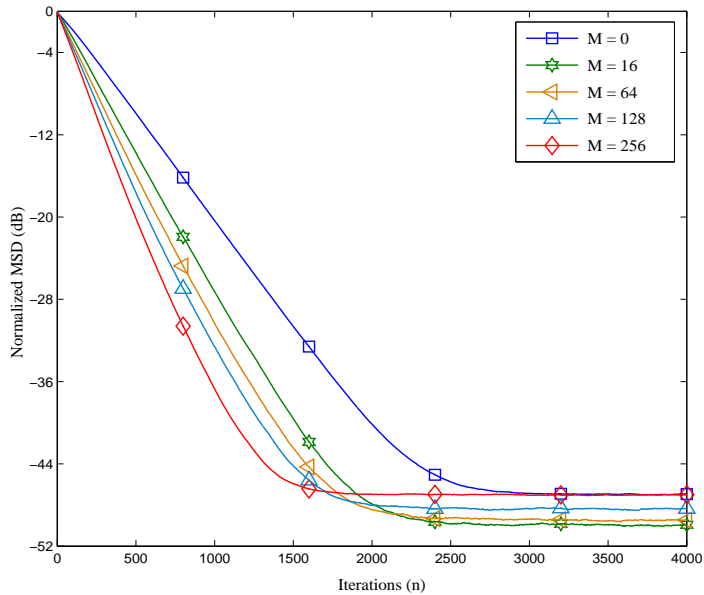}
\caption{Multi-task partial diffusion APA with projection order $=8$}
\label{the-label-for-cross-referencing}
\end{figure}
From the above results we can make the following observations:
\begin{itemize}
\item Multi-task partial diffusion APA exhibits a tradeoff between the communication load and the estimation performance.
\item AS expected, multi-task partial diffusion APA convergence rate was slower than multi-task full diffusion APA, however its steady state performance was better than the full multi-task diffusion APA. The partial diffusion during the adaptation step affect the convergence rate however improves the steady state performance. Theoretically,  we can observe the eigen value of $\textbf{F}$ is greater than the eigen value of $\textbf{F}_{M}$ i.e., $\lambda_{i}(\textbf{F}_{L}) \geq \lambda_{i}(\textbf{F}_{M}) $ where $M \leq L$.
\end{itemize}

\section{Conclusions}
In this paper, we presented the multi-task partial diffusion APA strategies which are suitable for multi-task networks that need less communication load. The proposed strategy is also robust against the correlated input conditions. The performance analysis of the proposed multi-task diffusion APA is presented in mean and mean square sense.

\end{document}